\documentclass[iop,revtex4]{emulateapj}
\usepackage[flushleft]{threeparttable}
\usepackage{graphics,graphicx,subfigure}
\usepackage{epstopdf}
\usepackage{epsfig}
\usepackage{helvet}
\usepackage[utf8]{inputenc}
\usepackage[T1]{fontenc}
\usepackage{longtable}
\usepackage{soul}
\usepackage{amsmath, amssymb, gensymb} 
\usepackage{natbib}
\usepackage{microtype}
\usepackage{multirow}
\usepackage{float}
\usepackage{morefloats}
\usepackage{hyperref}
\usepackage{lscape}
\usepackage{outlines}
\usepackage{color}
\usepackage{appendix}
\usepackage{xspace}
\def\mjybm {mJy\,beam$^{-1}$\xspace}
\def\ujybm {$\mu$Jy\,beam$^{-1}$\xspace}
\def\mjybmvert {$\left(\frac{\textrm{mJy}}{\textrm{beam}}\right)$\xspace}

\def\etal {\textit{et al.}\xspace}

\begin{document}

\slugcomment{
Submitted to PASP on 22 Apr 2020; accepted on 4 Jun 2020
}

\title{Characterizing the accuracy of ALMA linear-polarization mosaics}
\shorttitle{Characterizing the accuracy of ALMA linear-polarization mosaics}

\author{Charles L. H. Hull\altaffilmark{1,2,12}}
\author{Paulo C. Cortes\altaffilmark{3,2}}
\author{Valentin J. M. Le Gouellec\altaffilmark{4,5}}
\author{Josep M. Girart\altaffilmark{6,7}}
\author{Hiroshi Nagai\altaffilmark{8,9}}
\author{Kouichiro Nakanishi\altaffilmark{8,9}}
\author{Seiji Kameno\altaffilmark{1,2,9}}
\author{Edward B. Fomalont\altaffilmark{3,2}}
\author{Crystal L. Brogan\altaffilmark{3}}
\author{George A. Moellenbrock\altaffilmark{10}}
\author{Rosita Paladino\altaffilmark{11}}
\author{Eric Villard\altaffilmark{4,2}}

\altaffiltext{1}{National Astronomical Observatory of Japan, NAOJ Chile, Alonso de C\'ordova 3788, Office 61B, 7630422, Vitacura, Santiago, Chile}
\altaffiltext{2}{Joint ALMA Observatory, Alonso de C\'ordova 3107, Vitacura, Santiago, Chile}
\altaffiltext{3}{National Radio Astronomy Observatory, 520 Edgemont Road, Charlottesville, VA 22903, USA}
\altaffiltext{4}{European Southern Observatory, Alonso de C\'ordova 3107, Vitacura, Santiago, Chile}
\altaffiltext{5}{AIM, CEA, CNRS, Universit\'e Paris-Saclay, Universit\'e Paris Diderot, Sorbonne Paris Cit\'e, F-91191 Gif-sur-Yvette}
\altaffiltext{6}{Institut de Ci\`encies de l'Espai (ICE-CSIC), Campus UAB, Carrer de Can Magrans S/N, E-08193 Cerdanyola del Vall\`es, Catalonia}
\altaffiltext{7}{Institut d'Estudis Espacials de Catalunya, E-08030 Barcelona, Catalonia}
\altaffiltext{8}{National Astronomical Observatory of Japan, 2-21-1 Osawa, Mitaka, Tokyo 181-8588, Japan}
\altaffiltext{9}{The Graduate University for Advanced Studies, SOKENDAI, Osawa 2-21-1, Mitaka, Tokyo 181-8588, Japan}
\altaffiltext{10}{National Radio Astronomy
Observatory, P.O. Box O, Socorro, NM 87801, USA}
\altaffiltext{11}{INAF-Istituto di Radioastronomia, via P. Gobetti, 101, 40129, Bologna, Italy}
\altaffiltext{12}{NAOJ Fellow}

\shortauthors{Hull \etal}
\email{chat.hull@nao.ac.jp}

\begin{abstract}
We characterize the accuracy of linear-polarization mosaics made using the Atacama Large Millimeter/submillimeter Array (ALMA). First, we observed the bright, highly linearly polarized blazar 3C~279 at Bands 3, 5, 6, and 7 (3\,mm, 1.6\,mm, 1.3\,mm, and 0.87\,mm, respectively).  At each band, we measured the blazar's polarization on an 11$\times$11 grid of evenly-spaced offset pointings covering the full-width at half-maximum (FWHM) area of the primary beam.  After applying calibration solutions derived from the on-axis pointing of 3C~279 to all of the on- and off-axis data, we find that the residual polarization errors across the primary beam are similar at all frequencies: the residual errors in linear polarization fraction $P_\textrm{frac}$ and polarization position angle $\chi$ are $\lesssim$\,0.001 ($\lesssim$\,0.1\% of Stokes $I$) and $\lesssim$\,1$\degree$ near the center of the primary beam; the errors increase to $\sim$\,0.003--0.005 ($\sim$\,0.3--0.5\% of Stokes $I$) and $\sim$\,1--5$\degree$ near the FWHM as a result of the asymmetric beam patterns in the (linearly polarized) $Q$ and $U$ maps.  We see the expected double-lobed ``beam squint'' pattern in the circular polarization (Stokes $V$) maps.  Second, to test the polarization accuracy in a typical ALMA project, we performed observations of continuum linear polarization toward the Kleinmann-Low nebula in Orion (Orion-KL) using several mosaic patterns at Bands 3 and 6.  We show that after mosaicking, the residual off-axis errors decrease as a result of overlapping multiple pointings.  Finally, we compare the ALMA mosaics with an archival 1.3\,mm CARMA polarization mosaic of Orion-KL and find good consistency in the polarization patterns.
\\
\end{abstract}


\section{Introduction}
\label{sec:intro} 

When an astronomical source is not observed at the pointing center, off-axis errors in linear and circular polarization will affect the resulting observations of polarized emission. This is true for all telescopes.  In the case of telescopes with on-axis receiver feeds, off-axis errors in the linear polarization appear.  In those with receiver feeds whose axes are offset with respect to the axis of the reflector, off-axis errors in both linear and circular polarization become evident \citep{Chu1973}; and indeed, the vast majority of telescope with multiple receiver bands fall into this latter category, including, e.g., the Atacama Large Millimeter/submillimeter Array (ALMA) and the Karl G. Jansky Very Large Array (VLA).  

In addition to the traditional method of full-polarization holographic imaging \citep[e.g.,][]{Harp2011, Perley2016, Jagannathan2017}, another method of characterizing the wide-field polarization errors across the primary beam of an interferometer is to use the full array to perform a grid of observations, where all antennas simultaneously observe a polarized point source (e.g., a quasar or blazar) in many offset positions.  Here we report results of 11$\times$11 observations by ALMA at Bands 3, 5, 6, and 7 (3\,mm, 1.6\,mm, 1.3\,mm, and 0.87\,mm, respectively) toward the highly linearly polarized blazar 3C~279.\footnote{Initial tests using this method at Bands 3, 6, and 7 were reported by Hiroshi Nagai in ALMA System Verification Report SYS \# 225: Off-axis cross polarization (SYSE-88.00.00.00-0037-B-REP).}  Since the wide-field polarization response of an antenna manifests itself in the antenna's frame of reference, in all tests the offset pointings were evenly spaced in the azimuth-elevation (Azimuth,\,Elevation) frame, not in the (RA,\,DEC) frame like a typical observation.

\begin{figure} [hbt!]
\centering
\includegraphics[scale=0.31, clip, trim=0cm 0cm 9cm 0cm]{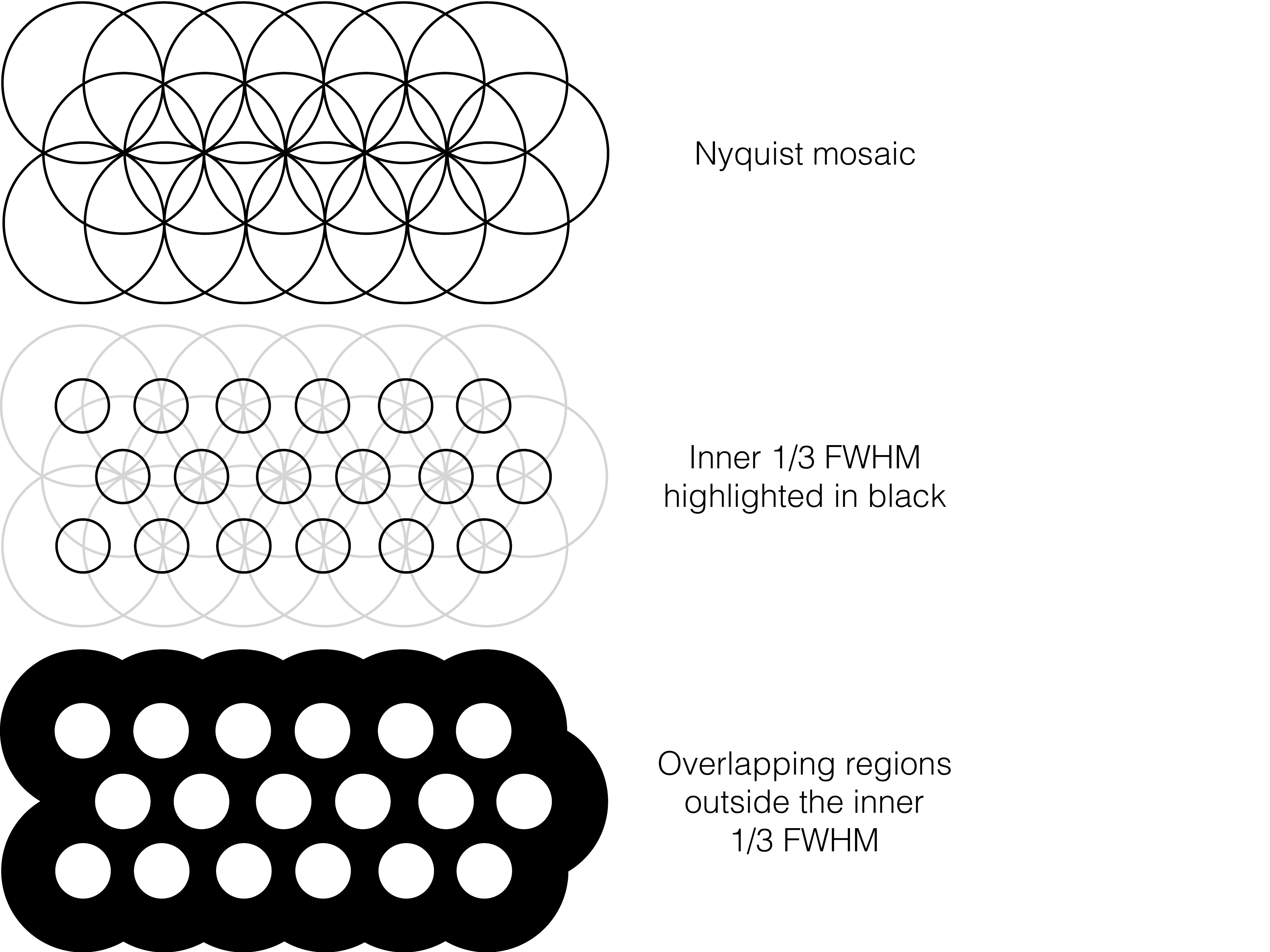}
\smallskip
\caption{\small \textit{Top:} An illustration of a standard Nyquist mosaic, where the diameter of each circle indicates the FWHM of each pointing.  Each pointing is centered at the FWHM of neighboring pointings.  \textit{Middle:} small black circles indicate the inner $\frac{1}{3}$\,FWHM, where the off-axis polarization errors are minimal.  \textit{Bottom:} the black shaded region indicates the areas of the mosaic comprising emission that falls outside of the $\frac{1}{3}$\,FWHM of any pointing.
}
\label{fig:mosaic_packing}
\bigskip
\end{figure}

A primary reason for characterizing the wide-field polarization performance of ALMA antennas is to understand the effect that wide-field polarization errors have in a polarization mosaic, where many pointings are stitched together.  Performing wide-field polarization science using an image made from a single ALMA pointing is not advisable because the polarization performance far from the center of the primary beam is sub-optimal due to residual off-axis errors.  However, mosaicking an image alleviates this problem, because many pointings are combined, and the on-axis emission (i.e., emission located at or very near the pointing center) in any giving pointing is more heavily weighted than the off-axis emission.  On the other hand, in contrast to a non-mosaicked, on-axis, single-pointing observation, a mosaic contains \textit{some} off-axis emission from one or more adjacent pointings in every location in the image: see, for example, the black shaded region in Figure \ref{fig:mosaic_packing}, which indicates the regions of a standard Nyquist mosaic that do not coincide with the inner $\frac{1}{3}$\,FWHM of any given pointing.\footnote{At the time of publication, proposals for single-pointing polarization observations with ALMA may only target sources whose emission falls within the inner $\frac{1}{3}$\,FWHM at the requested band, which is where the off-axis polarization errors are minimal.}  This emission from off-axis regions of multiple neighboring pointings could corrupt the final image, and thus here we also characterize the error in a mosaicked image by analyzing linear-polarization observations of an extended, highly linearly polarized source with ALMA at Bands 3 and 6 (3\,mm and 1.3\,mm, respectively).  

Our target is the Kleinmann-Low Nebula in Orion (Orion-KL), which lies in the OMC-1 region at the center of the ``integral-shaped filament,'' a filamentary cloud with a length of  $\gtrsim$\,7\,pc that lies in the northern portion of the Orion A star-forming cloud \citep{Johnstone1999}.  Our observations of Orion-KL comprise a mosaic centered on the iconic Source I \citep[see, e.g.,][and references therein]{Plambeck2009}, and extending NNE and SSW to cover the filamentary structure, including the ``Northern Ridge'' \citep{Johnstone1999, Hull2014} that lies $\sim$\,25$\arcsec$ to the NE of Source I.  Later in this work, we compare our ALMA mosaics with a similar 1.3\,mm linear-polarization mosaic of Orion-KL performed by \citet{Hull2014} using the Combined Array for Research in Millimeter-wave Astronomy (CARMA).

Single-dish polarization observations using the POL2 camera on the James Clerk Maxwell Telescope (JCMT) by \citet{Pattle2017} yielded a high-sensitivity image of the magnetic field in Orion A, which is roughly perpendicular to the filament's long axis, as frequently seen in star-forming clouds whose magnetic field is thought to be dynamically important \citep[e.g.,][]{Fissel2016, PlanckXXXV}.  These JCMT results, along with the results that we present here and in a companion paper \citep{Cortes2020} are recent contributions to a large body of literature comprising single-dish and interferometric studies of millimeter, submillimeter, and far-infrared polarization toward the Orion-KL region and its environs.  The first detection of polarization toward Orion-KL in this wavelength regime was made by a balloon-borne polarimeter \citep{Cudlip1982}; numerous later detections used polarimeters on the Kuiper Airborne Observatory (KAO), the Berkeley-Illinois-Maryland Association (BIMA) array, the JCMT, the Caltech Submillimeter Observatory (CSO),  the Submillimeter Array (SMA), CARMA, the Stratospheric Observatory for Infrared Astronomy (SOFIA), and the IRAM 30\,m telescope \citep{Hildebrand1984, Novak1989, Rao1998, Schleuning1998, Plambeck2003, Girart2004, Houde2004, Matthews2009, Tang2010, Hull2014, Ritacco2017, Chuss2019}.  See \citet{Pattle2017} for a more exhaustive list of references.

Below we begin with a description of the observations, calibration, and imaging procedures for the 11$\times$11 observations of 3C~279 (\S\,\ref{sec:obs_11x11}), after which we discuss the results from these tests (\S\,\ref{sec:results_11x11}).  Similarly, we then present the mosaicked polarization observations toward Orion-KL (\S\,\ref{sec:obs_orion}), and the results from those mosaic tests (\S\,\ref{sec:results_orion}); the latter section includes a comparison between ALMA and CARMA observations toward Orion-KL (\S\,\ref{sec:carma}).  We end by offering our conclusions (\S\,\ref{sec:conc}). 

All of the data we present here, and in a companion paper by \citet{Cortes2020}, were collected as part of the ALMA Extension and Optimization of Capabilities (EOC) program.  The EOC program is based at the Joint ALMA Observatory (JAO) in Santiago, Chile---with collaboration from scientists both at the ALMA Regional Centers (ARCs) as well as at external institutions---and is focused on using science-quality data to test, verify, and open new observing modes at ALMA.

\begin{table*}[tbh!]
\centering
\normalsize
\caption{\normalsize Observational details (11$\times$11)}
\begin{tabular}{ccccl}
\hline \noalign {\smallskip}
Band & Obs. date & Source & $N_\textrm{ant}$ & UID  \\
     &      (UTC)       &        &                 &     \vspace{0.05in} \\
\hline \noalign {\smallskip}
3 & 2015 Mar 06 & 3C~279 & 31 & A002\_X9b98ec\_X94d \\ 
   &                     &  & & A002\_X9b98ec\_Xb98 \\
   &                     &  & & A002\_X9b98ec\_Xde3 \smallskip \\
5 & 2016 Nov 10 & 3C~279 & 9 & A002\_Xba6edb\_X31db \\
   &                     &  & & A002\_Xba6edb\_X3988 \\
   &                     &  & & A002\_Xba6edb\_X3d57 \smallskip \\
6 & 2015 May 08 & 3C~279 & 32 & A002\_Xa018c4\_X743 \\
   &                     &  & & A002\_Xa018c4\_Xa3d \\
   &                     &  & & A002\_Xa018c4\_Xdc7 \smallskip \\
7 & 2015 May 09 & 3C~279 & 35 & A002\_Xa018c4\_X3b92 \\
   &                     &  & & A002\_Xa018c4\_X3de3 \\
   &                     &  & & A002\_Xa018c4\_X402a
\end{tabular}
\smallskip
\tablecomments{\small 11$\times$11 observations toward the highly linearly polarized blazar 3C~279 (also known as J1256--0547).  
$N_\textrm{ant}$ is the number of antennas in the observation.
The UIDs each refer to an individual execution/observation.  
Note that while the Band 5 observations used fewer antennas than the other observations, this did not limit our results, as the observations had a sufficient signal-to-noise ratio (SNR).}
\label{table:data_11x11}
\bigskip
\end{table*}

\section{11$\times$11 observations of 3C~279}
\label{sec:obs_11x11} 

\begin{figure} [hbt!]
\centering
\includegraphics[scale=0.33, clip, trim=0.5cm 0.5cm 9cm 0.5cm]{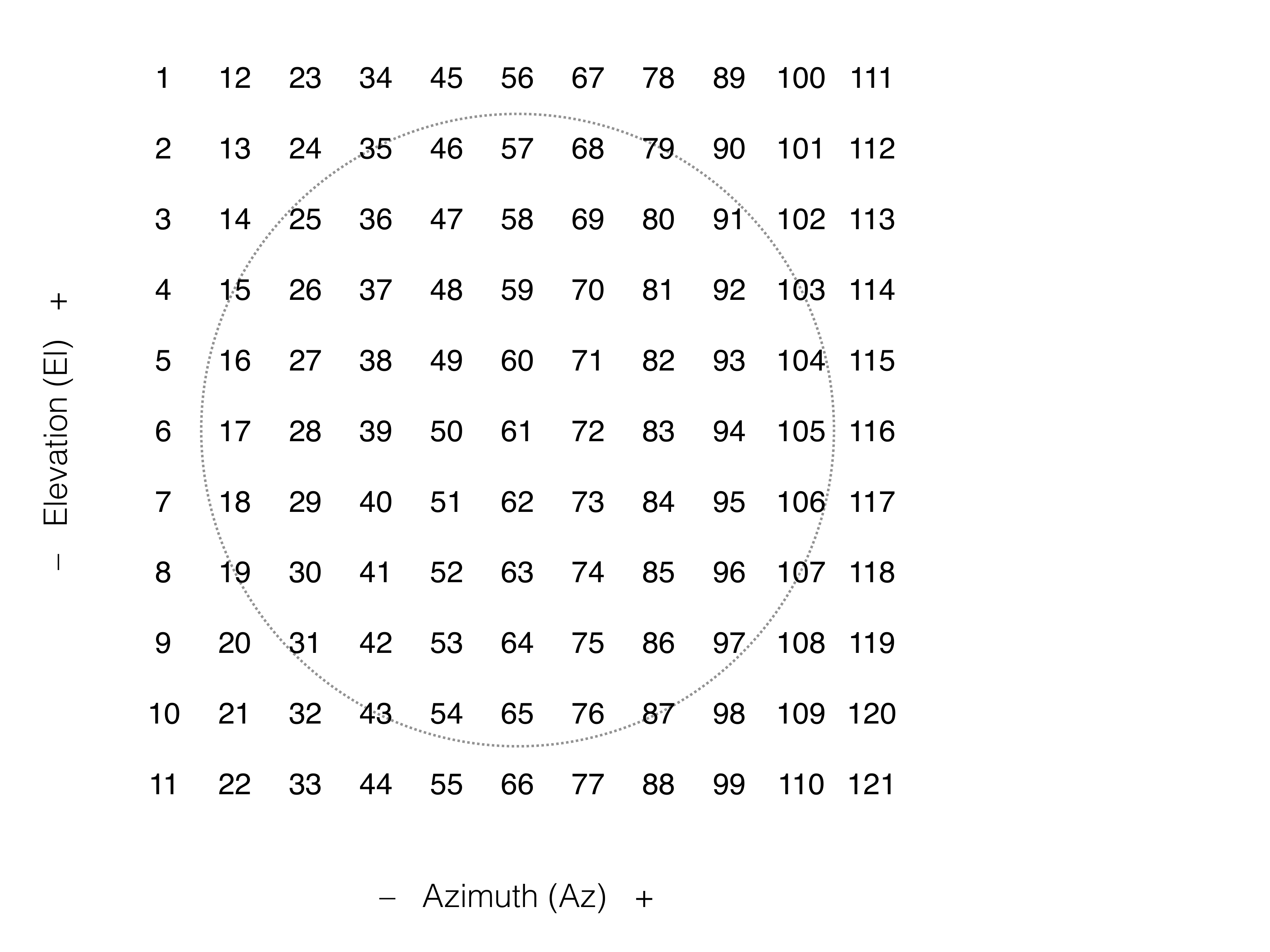}
\caption{\small 11$\times$11 pointing pattern.  The dotted gray circle indicates the FWHM of the primary beam.
}
\label{fig:11x11_pattern}
\bigskip
\end{figure}

Our first set of observations comprise 11$\times$11 grids of offset pointings toward 3C~279 at Bands 3, 5, 6 and 7, the goal of which is to characterize the off-axis polarization errors across the field of view of single-pointing ALMA observations at several ALMA bands.  The 11$\times$11 grid pointings are evenly spaced in (Azimuth,\,Elevation), and extend out to approximately the FWHM of the primary beam at each observing frequency.\footnote{The actual size of the ALMA primary beam is slightly larger than the standard assumed beam size of $1.22\,\lambda / D,$ where $\lambda$ is the observing wavelength and $D$ is the antenna diameter.  For more details, see \url{https://safe.nrao.edu/wiki/pub/ALMA/NAASC/Memo114Appendices/sdimagingEDM.pdf}.}  Figure \ref{fig:11x11_pattern} shows the pattern of the 121 pointings.  Each set of observations of 3C~279 includes three separate executions spaced by $\sim$\,1\,hr, which yields sufficient parallactic angle coverage to allow us to solve for the leakage terms ($D$-terms) and the cross-hand phase ($XY$-phase); we discuss the derivation of both of these quantities later in this section.  We list the observation dates and data identification codes (or UIDs, each corresponding to an individual ALMA execution/observation) in Table \ref{table:data_11x11}.  We chose to observe 3C~279 because, at the time of observation in 2015/2016, it was both extremely bright and had an extremely high linear polarization fraction.  All observations were taken using the standard correlator setup for continuum polarization observations, which includes four 1.875\,GHz-wide spectral windows with 64 channels each. See Table \ref{table:obs_11x11} for the on-axis results, and for the observational setup at each band.    

\begin{table*}[tbh!]
\centering
\normalsize
\caption{\normalsize Observational setup and on-axis results (11$\times$11)}
\begin{tabular}{ccccccccc}
\hline \noalign {\smallskip}
Band & $\nu$ & $\theta$ & Source & $I$ & $Q$ & $U$ & $P_\textrm{frac}$ & $\chi$  \\
& (GHz) & & & & & & & (deg) \vspace{0.05in} \\
\hline \noalign {\smallskip}
3 & \phantom{3}97.479 & $3\farcs31 \times 2\farcs49$ & 3C~279 & 1 & 0.0404 & 0.1074 & 0.115 & 34.7 \\ 
5 & 183.261 & $1\farcs03 \times 0\farcs61$ & 3C~279 & 1 & --0.0005\phantom{2} & 0.1095 & 0.110 & \phantom{0}45.1$^a$ \\
6 & 233.000 & $1\farcs15 \times 0\farcs88$ & 3C~279 & 1 & 0.0398 & 0.1142 & 0.121 & 35.4 \\
7 & 343.479 & $0\farcs82 \times 0\farcs55$ & 3C~279 & 1 & 0.0373 & 0.1158 & 0.122 & 36.1
\end{tabular}
\smallskip
\tablecomments{\small 
Results from the 11$\times$11 observations, listed in the same order as in Table \ref{table:data_11x11}.  
$\nu$ is the average frequency of the observations.
$\theta$ is the synthesized beam (resolution element) of the images.
All fluxes are peak flux values, and are scaled relative to a normalized peak Stokes $I$ flux of 1.  
The absolute Stokes $I$ flux densities of 3C~279 were $\sim$\,14.7\,Jy (Band 3, 2015 Mar), $\sim$\,6.8\,Jy (Band 5, 2016 Nov, interpolated using the available Band 3 and Band 7 fluxes from the ALMA Calibrator Source Catalogue: \url{https://almascience.eso.org/sc}), $\sim$\,7.5\,Jy (Band 6, 2015 May), and $\sim$\,5.6\,Jy (Band 7, 2015 May). \smallskip
\\
$^a$ The change in polarization angle of 3C~279 from $\sim$\,35$\degree$ in Mar--May 2015 (when the Band 3, 6, 7 data were taken) to $\sim$\,45$\degree$ in Nov 2016 (when the Band 5 data were taken) can be seen in all frequency bands; see \url{http://www.alma.cl/~skameno/AMAPOLA}.
}
\label{table:obs_11x11}
\bigskip
\end{table*}

Note that, as no other calibrators besides 3C~279 were observed in the 11$\times$11 tests, we perform every step of the reduction process using 3C~279.  We first split out the scans where 3C~279 was located at the pointing center (i.e., the central point of the 11$\times$11 grid) and concatenate them into a separate dataset; we use these on-axis data to derive the on-axis polarization calibration via the current ALMA polarization calibration scheme.  For further details about standard, on-axis reduction of ALMA polarization data, see \citet{Nagai2016} as well as the 3C~286 Polarization CASA Guide.\footnote{\,3C 286 Polarization CASA Guide: \url{https://casaguides.nrao.edu/index.php?title=3C286_Polarization}}    

We first perform an initial phase-versus-time gain calibration, which we apply on-the-fly when solving for the bandpass solution.  After solving for the bandpass, we perform gain calibration (amplitude and phase), assuming an initial unpolarized source model for 3C~279.  
Since ALMA has crossed-linear feeds, the linear polarization properties of the source manifest themselves in the gain amplitude solutions when an initial, unpolarized model is used.  To make a first estimate of the polarization of 3C~279, we use the task \texttt{qufromgain} from \texttt{almapolhelpers.py}.  Next, we solve for the cross-hand delay using the scans where the $XX$/$YY$ polarization ratio is changing the fastest (this can be seen by viewing the gain-amplitude calibration table).  We then solve for the $XY$-phase, which is the phase difference between the $X$ and the $Y$ polarizations of the reference antenna.  This solution can have angle ambiguities, which we resolve by using the task \texttt{xyamb}; this task uses both the initial $XY$-phase solution as well as the previously derived initial model of the source polarization to solve for the final polarization model of 3C~279.  With the correct model of 3C~279 in hand,\footnote{Note that, at the time of publication, the polarization calibration procedure with ALMA includes the assumption that the calibrator Stokes $V = 0$.  This may not necessarily be the case, as quasars/blazars may have faint intrinsic circular polarization at millimeter wavelengths at levels of $\pm$ a few $\times$ 0.1\% of the Stokes $I$ value \citep[e.g.,][]{Thum2018}.} we then re-derive the gains (amplitude and phase) using the model in order to remove the polarization response from the gain amplitudes.  Finally, we derive the $D$-terms.  After doing so, we apply all of the calibration tables (the bandpass; the modified gains after removing the source polarization; the cross-hand delay; the $XY$-phase; and the $D$-terms) to our on-axis dataset.  

After applying the standard on-axis calibration to all of the on- and off-axis data, we perform one round of phase-only self-calibration (selfcal) to correct the residual phase errors as a function of time for both the on-axis data and each off-axis pointing.  With the exception of the selfcal phase solutions, all calibration tables that we apply to the off-axis data are solved for using \textit{on-axis data only}---this includes $D$-terms and $XY$-phase.  We do this to mimic the standard calibration method of ALMA polarization data.

After completing the data reduction process, we produce $I$,\,$Q$,\,$U$,\,$V$ images using the task \texttt{clean} from CASA version 5.6.1 with standard imaging parameters, including a Briggs visibility weighting of \texttt{robust}\,=\,2.0 (i.e., natural weighting).  After making the images, we extract the $I$,\,$Q$,\,$U$,\,$V$ peak values using the CASA task \texttt{imfit}; these are the values that we analyze and report in Section \ref{sec:results_11x11}.  Finally,  we use these Stokes maps to produce maps of the linearly polarized intensity $P$, linear polarization fraction $P_\textrm{frac}$, and linear polarization angle $\chi$: 

\begin{equation}
P = \sqrt{Q^2 + U^2}
\end{equation}
\begin{equation}
P_{\textrm{frac}} = \frac{P}{I}
\end{equation}
\begin{equation}
\chi = \frac{1}{2}\, \textrm{arctan2}\,{\left(\frac{U}{Q}\right)} \label{eqn:pa} \,\,.
\end{equation}




The dynamic range (peak flux value divided by the rms noise value) in the Stokes $I$ continuum maps ranges from $\sim$\,10,000--100,000, depending on the observing frequency (lowest dynamic range at Band 7, highest at Band 3).\footnote{Note that while we did perform one round of phase-only selfcal to improve the images, the improvements in the dynamic range of the Stokes $I$ images were modest: between 10--30\% for Bands 3, 6, and 7, and a factor of $\sim$\,2 for Band 5.}  The dynamic range in the polarized intensity ($P$) maps is always a factor of $\sim$\,10 lower, since 3C~279 had a polarization fraction of $\sim$\,10\% in all of our observations (see Table \ref{table:obs_11x11}). Nevertheless, the dynamic range of our $P$ maps is always >\,1000; in a region where the dynamic range of $P$ is 1000, the statistical error of the polarization angle ($\sigma_\chi = 0.5\,\sigma_P / P$) is $\sim$\,0.03$\degree$, which is much smaller than the on-axis uncertainty of $\sim$\,0.4$\degree$ in the polarization angle calculated by \citet{Nagai2016} (\citeauthor{Nagai2016} note that the exact value depends on the number of antennas used in each observation).  Therefore, changes in the image properties of the $I,$ $Q,$ and $U$ Stokes parameters as a function of offset pointing are not caused by sensitivity limits, but rather by systematic polarization variations across the primary beam.

Since the time when these data were obtained 4--5 years ago, there have been no major changes in the receiver optics.  The sub-reflectors in each antenna also remain un-tilted.  A number of ALMA 12\,m antennas were found to be astigmatic, which requires a corrective scheme to be applied to the surface of the dishes; however, as of the time of publication, this scheme has not yet been implemented.  If the antenna astigmatism introduces wide-field polarization errors, then the errors reported in this work should improve once the correction scheme is applied.  In the future, ALMA is planning new receivers, which will potentially change the optics (e.g., additional external mirrors) and will also use the tilting capabilities of the sub-reflector to improve the alignment with the receiver feeds. These changes will most likely improve the errors that we report here, but future tests will be needed to confirm this.

\section{Results: 11$\times$11 observations of 3C~279}
\label{sec:results_11x11} 

\subsection{Error in $Q$, $U$, $P_\textrm{frac}$, and $\chi$}
\label{sec:11x11_errors}

\begin{figure*} [hbt!]
\centering
\includegraphics[scale=0.58, clip, trim=0.9cm 0cm 0cm 0.9cm]{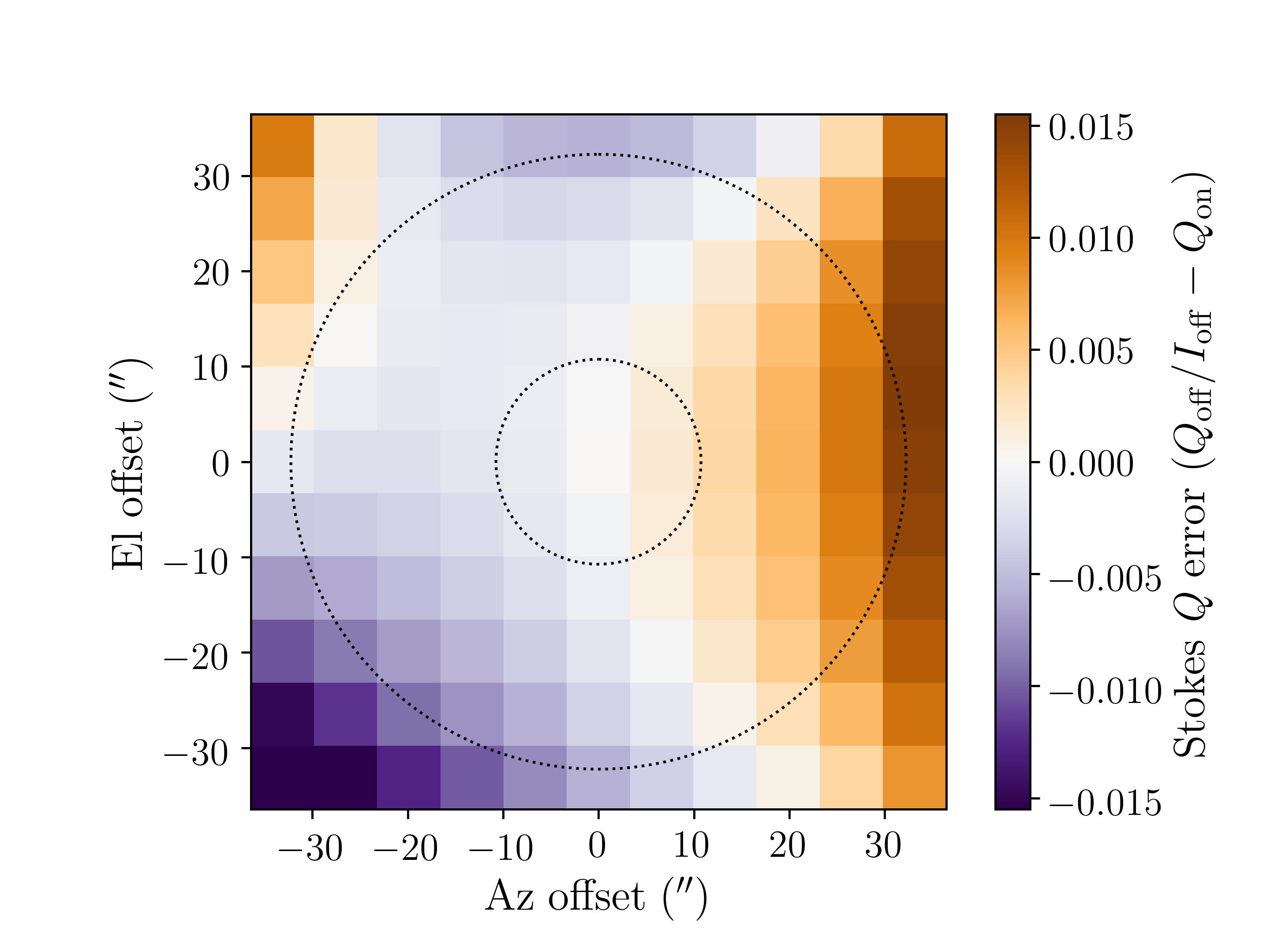}
\includegraphics[scale=0.58, clip, trim=0.9cm 0cm 0cm 0.9cm]{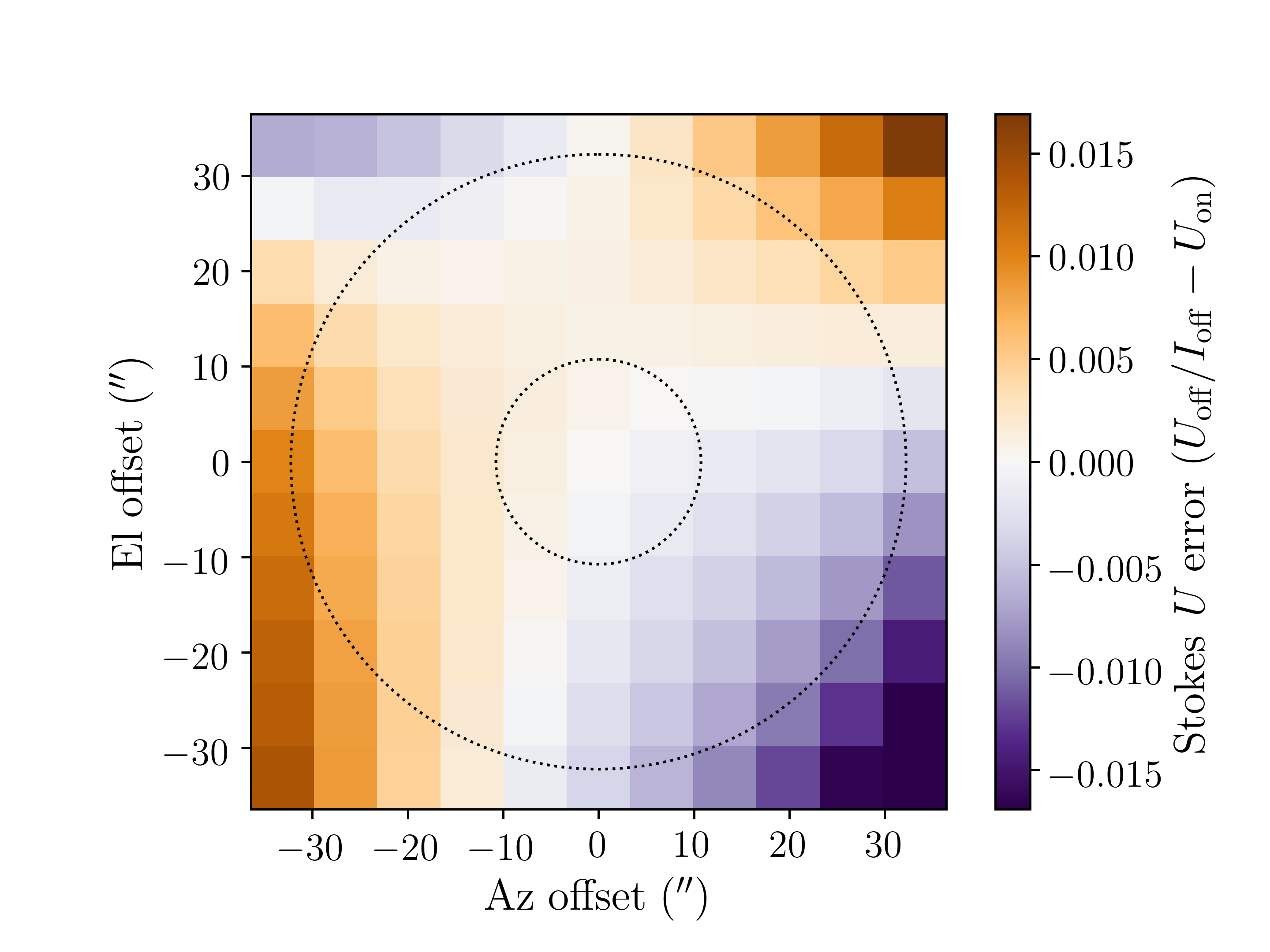}
\includegraphics[scale=0.58, clip, trim=0.9cm 0cm 0cm 0.9cm]{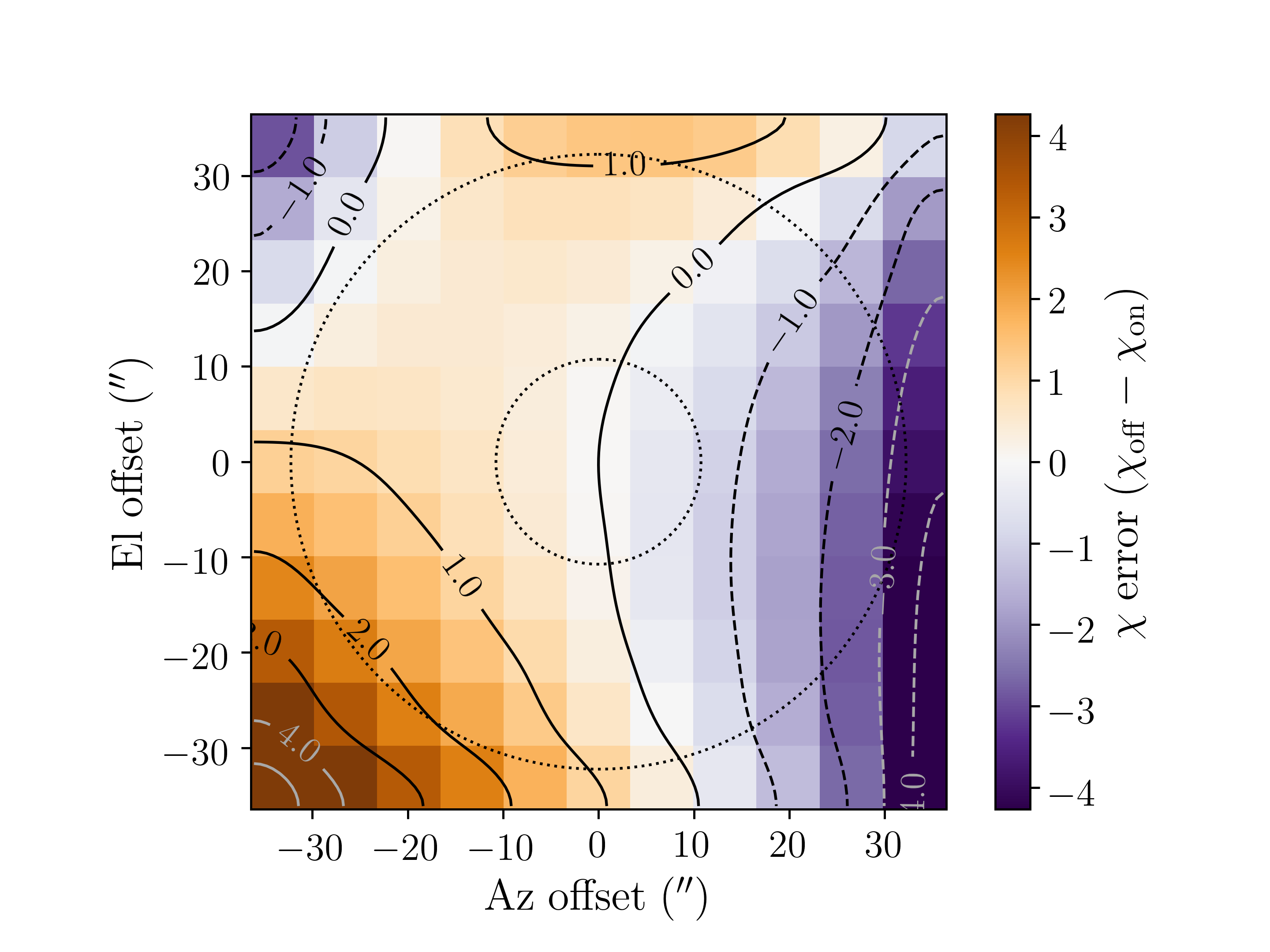}
\includegraphics[scale=0.58, clip, trim=0.9cm 0cm 0cm 0.9cm]{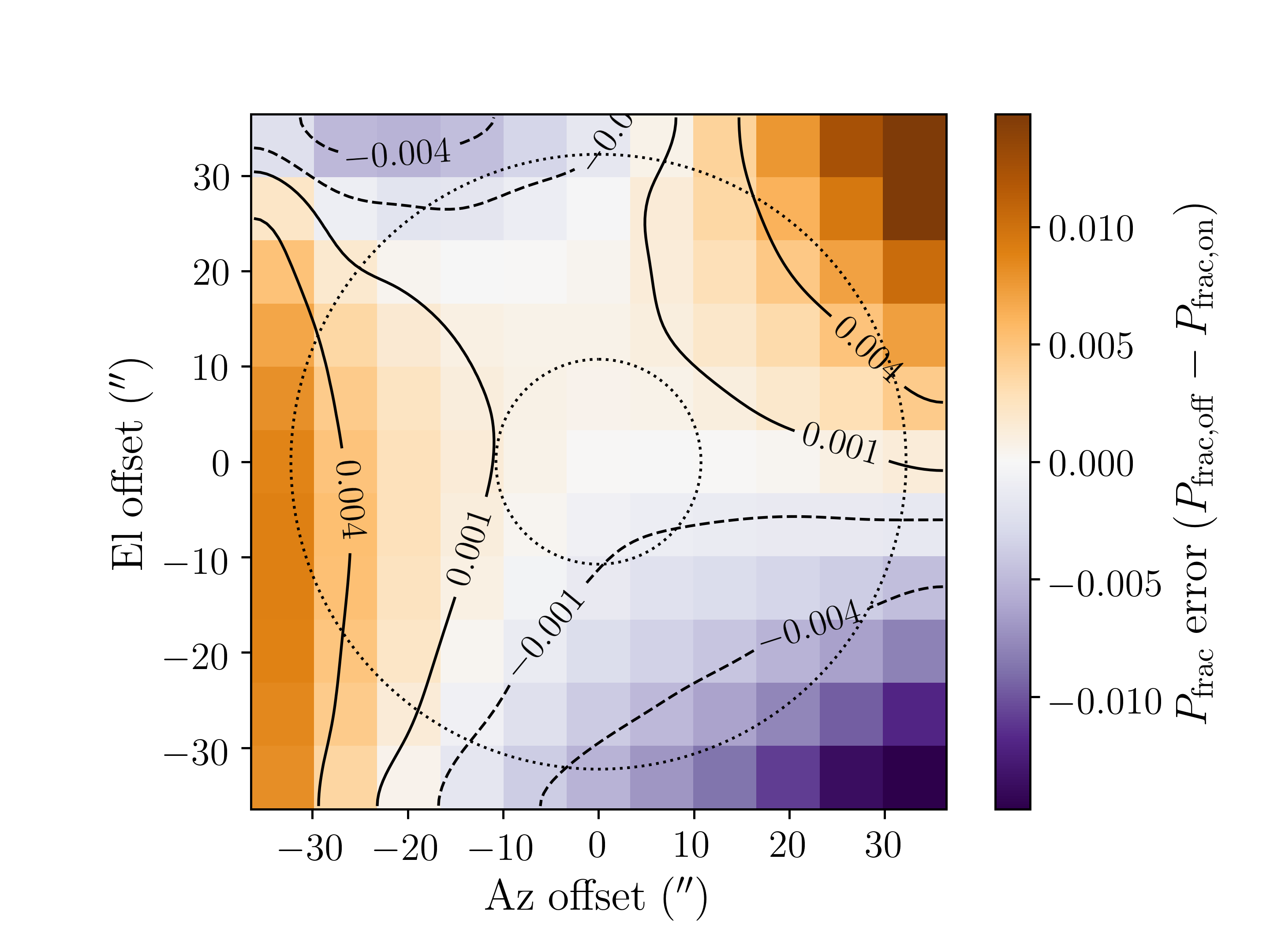}
\caption{\small Band 3 error maps in Stokes $Q,$ $U,$ position angle $\chi$, and polarization fraction $P_\textrm{frac}$. 
We apply on-axis calibration (including $D$-terms) to all off-axis positions.
The outer dotted line is the FWHM of the primary beam; the inner dotted line is the $\frac{1}{3}$\,FWHM level.  The primary beam response was removed from the $Q$ and $U$ error maps by dividing by the Stokes $I$ value in each pixel.
In all cases we subtract the central value from all of the off-axis pixels. 
The errors in $P_\textrm{frac}$ and $\chi$ shown in this figure (and similar figures for other bands) comprise the main results of the 11$\times$11 tests.}
\label{fig:errors_B3-3}
\bigskip
\end{figure*}

In Figure \ref{fig:errors_B3-3} (Band 3), and in Appendix \ref{app:data} (Bands 5, 6, and 7), we show the errors across the primary beam in Stokes $Q$, Stokes $U$, $\chi$, and $P_\textrm{frac}$ that we derived from the observations toward the highly linearly polarized blazar 3C~279, which had $P_\textrm{frac} \approx 12\%$ at the time of observation.  Assuming a circular, azimuthally symmetric primary beam in Stokes $I$, we can remove the primary beam response (i.e., the sensitivity fall-off with radius) in the Stokes $Q$ and $U$ images by dividing the off-axis Stokes $Q_\textrm{off}$ and $U_\textrm{off}$ values by the value of $I_\textrm{off}$ in the corresponding pixels:

\begin{align}
    Q_\textrm{off,norm} &= Q_\textrm{off} / I_\textrm{off} \\
    U_\textrm{off,norm} &= U_\textrm{off} / I_\textrm{off}\,\,.
\end{align}

\noindent
We then subtract the on-axis Stokes $Q_\textrm{on}$ or $U_\textrm{on}$ value from the normalized value in each pixel, leaving the residual errors $\delta Q$ and $\delta U$:  

\begin{align}
    \delta Q &= Q_\textrm{off,norm} - Q_\textrm{on} \label{eqn:deltaQ} \\ 
    \delta U &= U_\textrm{off,norm} - U_\textrm{on} \label{eqn:deltaU}    \,\,.
\end{align}

We define the off-axis $P_\textrm{frac,off}$ to be

\begin{equation}
    P_\textrm{frac,off} = \frac{\sqrt{Q_\textrm{off}^2 + U_\textrm{off}^2}}{I_\textrm{off}}\,\,.
\end{equation}

\noindent
As all of our Stokes $I$ maps are normalized to $I_\textrm{on} = 1$, this simplifies the equations for $P_\textrm{frac}$:

\begin{align}
    P_\textrm{frac,on} &= \sqrt{Q_\textrm{on}^2 + U_\textrm{on}^2} \\
    P_\textrm{frac,off} &= \sqrt{Q_\textrm{off,norm}^2 + U_\textrm{off,norm}^2}\,\,.
\end{align}

\noindent
Finally, we calculate the error in the position angle $\delta \chi$ and in the polarization fraction $\delta P_\textrm{frac}$ as the differences in the on- versus off-axis pointings:

\begin{align}
    \delta \chi &= \chi_\textrm{off} - \chi_\textrm{on} \\
    \delta P_\textrm{frac} &= P_\textrm{frac,off} - P_\textrm{frac,on}\,\,.
\end{align}

\noindent
In the error maps of $\chi$ and $P_\textrm{frac}$, we subtract the central value from each pixel.

After analyzing the observations of the highly linearly polarized blazar 3C~279, we find that the systematic errors in the polarization fraction $\delta P_\textrm{frac}$ and in the position angle $\delta\chi$ are similar for all observing frequencies.  Within the inner $\frac{1}{3}$\,FWHM, the residual errors in $P_\textrm{frac}$ and $\chi$ are $\lesssim$\,0.001 ($\lesssim$\,0.1\% of Stokes $I$) and $\lesssim$\,1$\degree$, respectively.  Near the FWHM, the errors increase to $\sim$\,0.003--0.005 ($\sim$\,0.3--0.5\% of Stokes $I$) and $\sim$\,1--5$\degree$, respectively, as shown in Figures \ref{fig:errors_B3-3}, \ref{fig:errors_B5-1}, \ref{fig:errors_B6-1}, and \ref{fig:errors_B7-1}.

\subsection{$V$ beam shape (beam squint)}
\label{sec:squint}

\begin{figure*} [hbt!]
\centering
\includegraphics[scale=0.58, clip, trim=0.9cm 0cm 0cm 0.9cm]{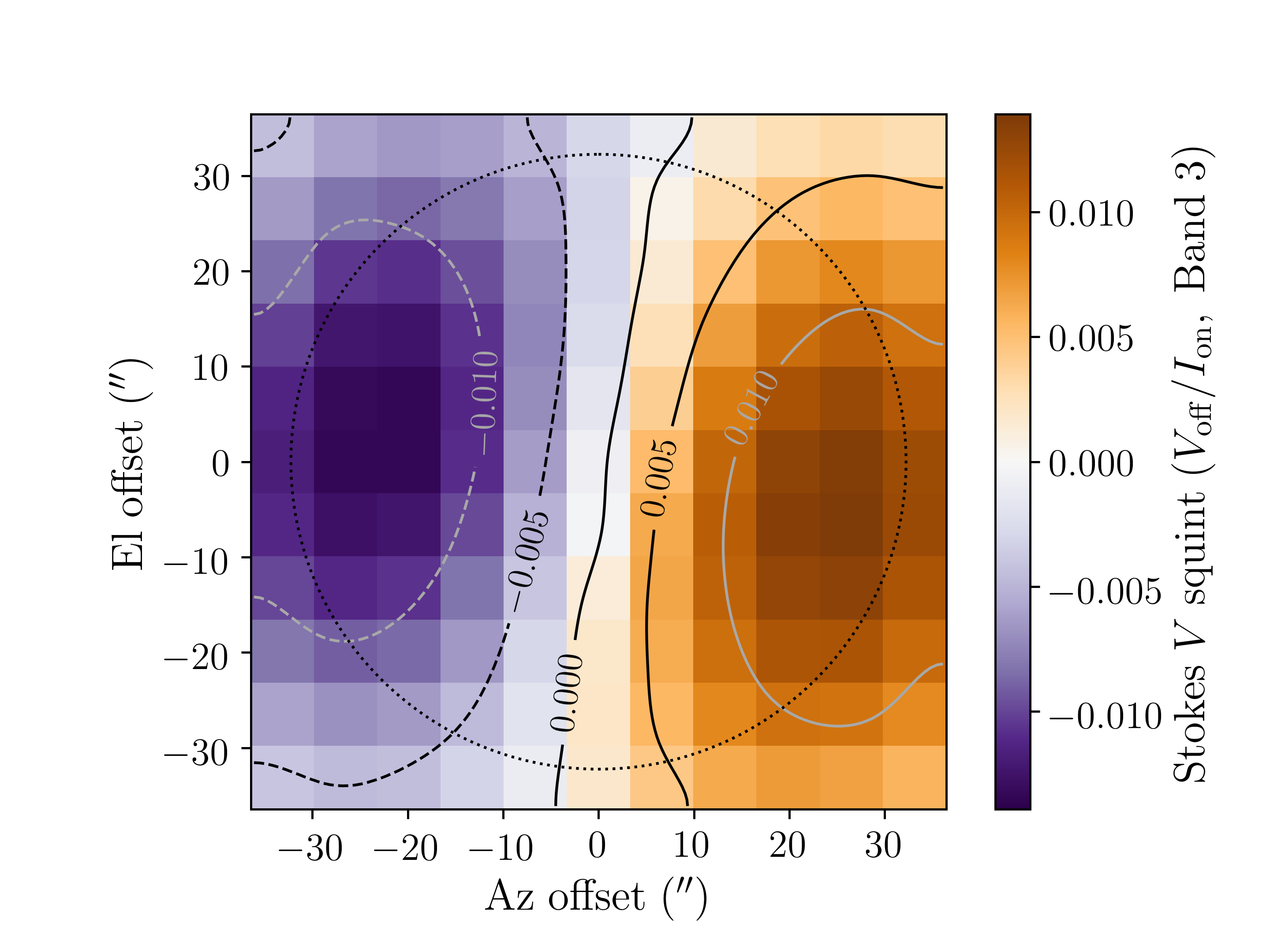}
\includegraphics[scale=0.58, clip, trim=0.9cm 0cm 0cm 0.9cm]{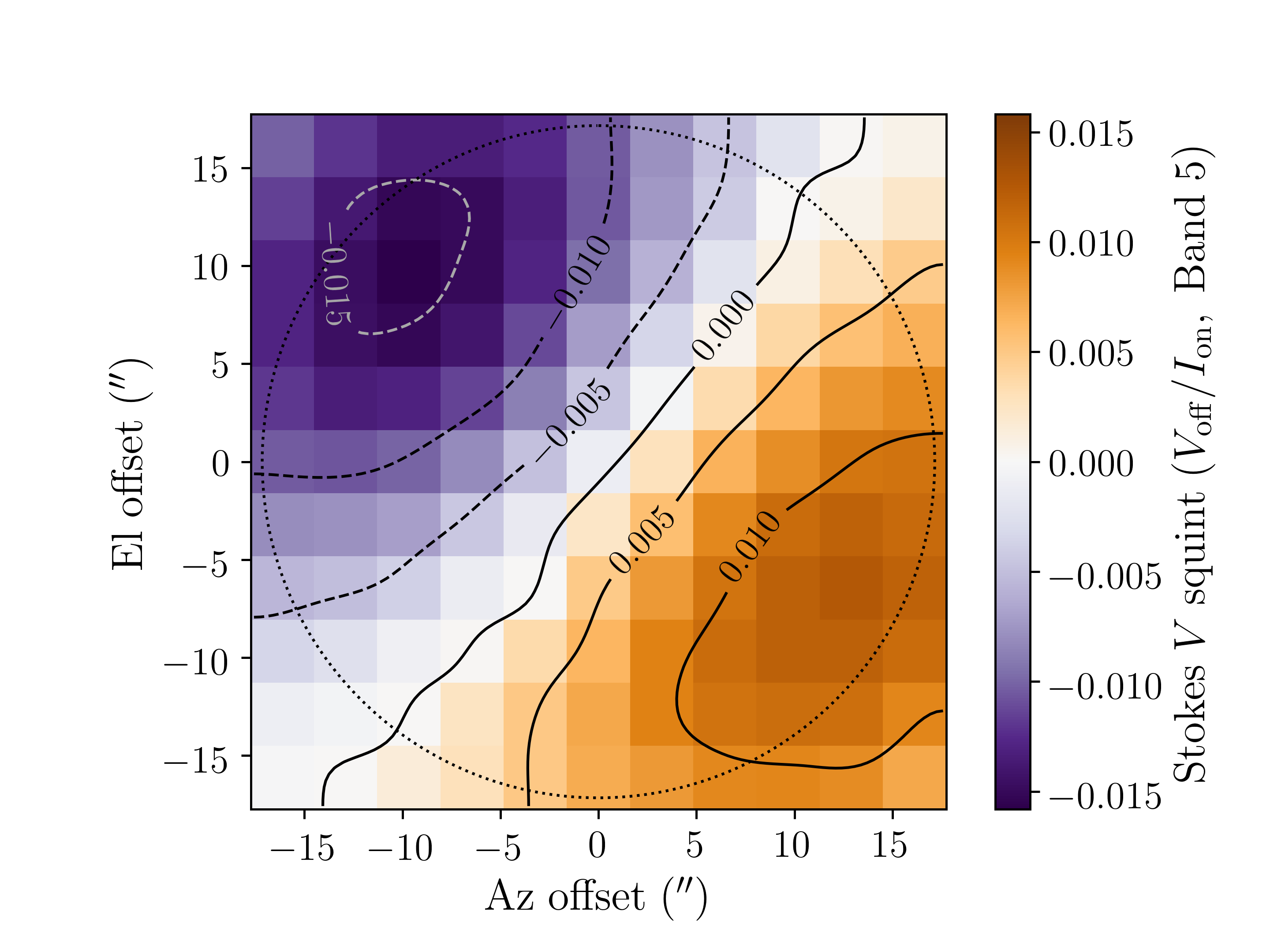} \\
\includegraphics[scale=0.58, clip, trim=0.9cm 0cm 0cm 0.9cm]{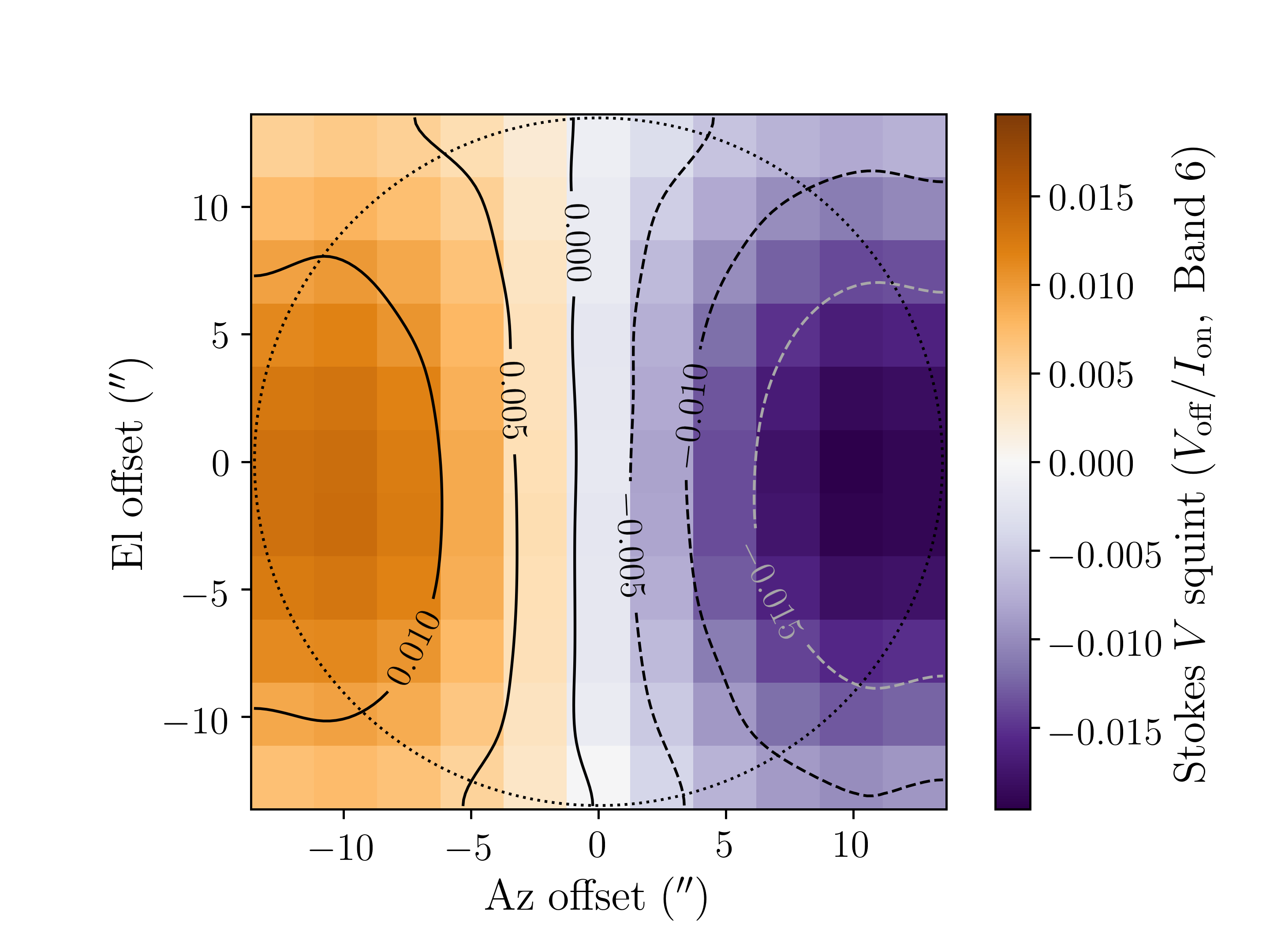}
\includegraphics[scale=0.58, clip, trim=0.9cm 0cm 0cm 0.9cm]{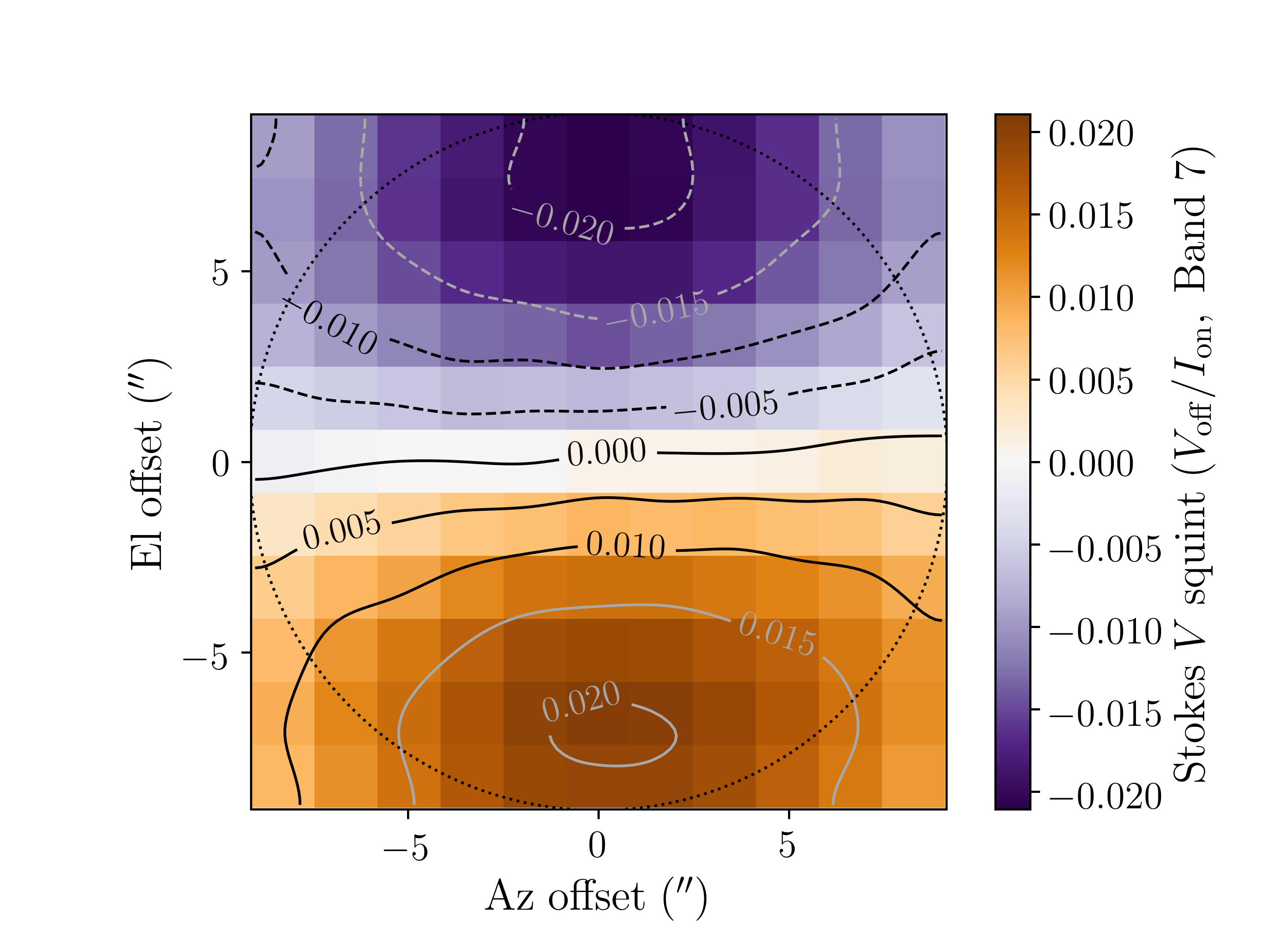}
\caption{\small Stokes $V$ squint profiles for Bands 3 (upper-left), 5 (upper-right), 6 (lower-left), and 7 (lower-right).  Colorscale shows $V/I_\textrm{on}$, where the on-axis value $I_\textrm{on}$ has been normalized to 1.  Note that the Stokes $V$ values have not been primary-beam corrected.}
\label{fig:data_V}
\bigskip
\end{figure*}

Each Stokes $V$ map exhibits a double-lobed ``beam squint'' pattern at the 1--2\% level of the on-axis Stokes $I$ value: see Figure \ref{fig:data_V} (note that the Stokes $V$ values have not been primary-beam corrected). Squint arises when a receiver is not aligned with the telescope's optical axis \citep{Chu1973, Adatia1975, Rudge1978}.\footnote{Note that even if a receiver were aligned with the optical axis of the telescope, beam squint would still arise if the receiver had two feed horns (like, e.g., the ALMA Bands 7, 9, and 10 receivers), as it is not possible for both horns to be aligned perfectly with the optical axis.}  This is the case for all ALMA receivers, which are installed in a single dewar along several concentric circles, all of which are offset from the optical axis \citep{Lamb2001}.  When dual-polarization receivers are offset, the response to left- and right-circular polarization (LCP and RCP, respectively) are slightly displaced from one another (this is true regardless of whether the receivers are crossed-linear like those at ALMA, or circular like, e.g., the 1.3\,mm receivers at CARMA).  Since Stokes $V$ $\equiv$ RCP -- LCP \citep{IEEE1997}, this offset in the circular-polarization response of the two separate receivers results in a double-lobed Stokes $V$ pattern. 

It is essential to correct for the known squint profile before attempting wide-field circular polarization observations (either mosaics or single pointings) with ALMA.  This is particularly true because unlike the more extended ``squash'' error pattern in the linearly polarized $Q$ and $U$ maps (see Section \ref{sec:squash}), the squint profile is compact, manifesting itself well within the FWHM at all bands.  Direction-dependent errors such as squint can be removed from VLA observations using the \texttt{awproject} keyword in the CASA task \texttt{tclean} \citep{Bhatnagar2008}; however, this has not yet been implemented for ALMA observations.  Efforts to implement full-polarization voltage-pattern corrections (i.e., primary beam models) at ALMA are underway (S. Bhatnagar et al., in preparation), and will allow us in the future to calibrate out the wide-field polarization errors that we analyze in this paper.

\begin{figure*}[tbh!]
\centering
\includegraphics[scale=0.45, clip, trim=0cm 0cm 1.37cm 0cm]{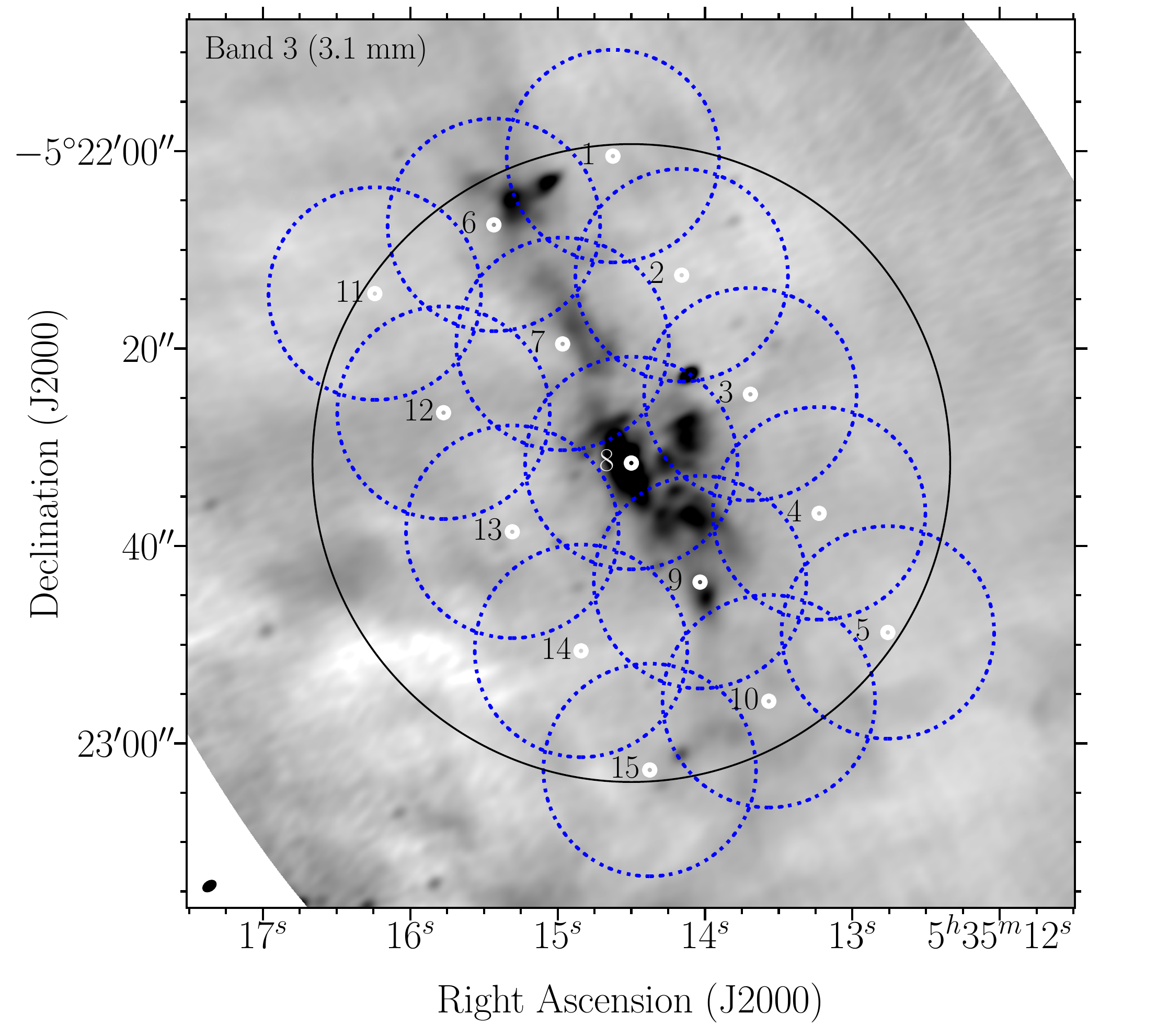}
\includegraphics[scale=0.45, clip, trim=3.4cm 0cm 1.3cm 0cm]{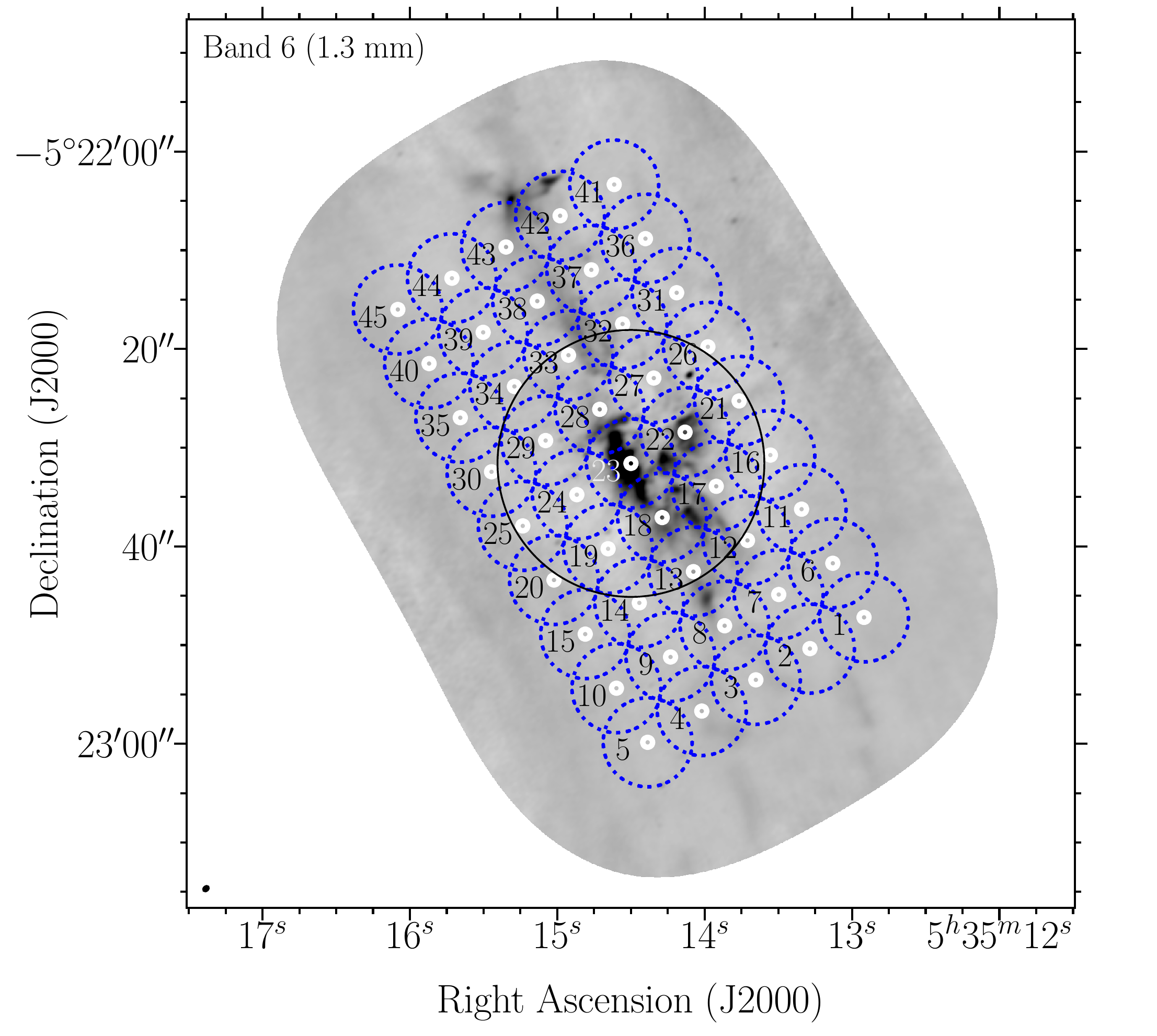}
\caption{ \small
Complete patterns for the Band 3 (left) and Band 6 (right) Orion-KL mosaics. The small white circles indicate the center of each field, the dotted blue circles correspond to the inner $\frac{1}{3}$\,FWHM region of the 12\,m antenna primary beam, and the solid black circle denotes the FWHM of the primary beam at the representative frequency of the observations.  The ellipses in the lower-left corners are the synthesized beams (resolution elements), which measure 1$\farcs$43\,$\times$\,0$\farcs$95 at Band 3 and 0$\farcs$65\,$\times$\,0$\farcs$48 at Band 6.
}
\bigskip
\label{fig:pattern}
\end{figure*}

\subsection{$Q$ and $U$ errors (beam squash)}
\label{sec:squash}

The maps of the off-axis errors in $Q$ and $U$ in some of the 11$\times$11 observations reveal hints of what is known as ``beam squash,''\footnote{The term ``beam squash'' was coined in \citet{Heiles2001c}.} which is a cloverleaf-like pattern in the Stokes $Q$ and $U$ error maps.  Like beam squint, beam squash is caused by the antenna optics.  Squash is the result of different beam widths of the orthogonal linear polarizations, which are in turn due to the varying projected geometries of the reflections over the parabolic reflector surface.  Beam squash is discussed in \citet{Napier1994, Napier1999}, and can be seen in maps made using the Arecibo Observatory \citep{Heiles2001c} and the Green Bank Telescope \citep[GBT;][]{Robishaw2018}.
Astroholography measurements show that the beam squash error patterns of the ALMA antennas have an angular extent of approximately twice that of the squint pattern (S. Bhatnagar et al., in preparation), and thus the extent of the 11$\times$11 observations is not sufficient to sample the entire pattern.  However, particularly in the Band 3 data (Figure \ref{fig:errors_B3-3}) and Band 7 data (Figure \ref{fig:errors_B7-1}), we begin to see the inner edge of the cloverleaf pattern in the error maps of $Q$ and $U$, with four lobes alternating between positive and negative.  We can also see the telltale 45$\degree$ rotation between the $Q$ and $U$ error patterns.  The squash error patterns of the ALMA antennas will be investigated more thoroughly in future studies focusing on full-polarization holographic mapping of the full ALMA primary beam (S. Bhatnagar et al., in preparation).

\section{Orion-KL mosaic observations}
\label{sec:obs_orion} 

We observed Orion-KL using the 12\,m array at Bands 3 and 6.  The data were taken in the standard sessions mode used for ALMA polarization observations \citep{THB}, and had the standard correlator setup for continuum polarization observations, which includes four 1.875\,GHz-wide spectral windows with 64 channels each.  The Band 3 and Band 6 sessions comprised three and two executions, respectively.  We list the details in Table \ref{table:data_orion}.  

\begin{table*}[tbh!]
\centering
\normalsize
\caption{\normalsize Observational details (Orion-KL)}
\begin{tabular}{cccccccl}
\hline \noalign {\smallskip}
Band & $\nu$ & Obs. date & Source & Config. & $\theta$ & $N_\textrm{ant}$ & UID  \\
     & (GHz) &    (UTC)       &        &         & &                  & \vspace{0.05in} \\
\hline \noalign {\smallskip}
3 & 97.479 & 2018 Mar 17 & Orion-KL & C-4 & $1\farcs43 \times 0\farcs95$ & 39 & A002\_Xca795f\_X50f \\
  & & 2018 Mar 18 &  &  & & & A002\_Xca795f\_Xcbd \\ 
  & & 2018 Mar 18 &  &  & & & A002\_Xca795f\_Xd0  \smallskip \\ 
6 & 233.000 & 2019 Apr 12 & Orion-KL & C-3 $\rightarrow$ C-4 & $0\farcs65 \times 0\farcs48$ & 43 & A002\_Xdab261\_X13a0a \\
  & & 2019 Apr 12 & &  & & & A002\_Xdab261\_X1448a \\
\end{tabular}
\smallskip
\tablecomments{\small Observations of Orion-KL.  
$\nu$ is the average frequency of the observations.
``Config.'' is the ALMA antenna configuration during which the observations were performed.
$\theta$ is the synthesized beam (resolution element) of the images.
$N_\textrm{ant}$ is the number of antennas in the observation.
The UIDs each refer to an individual execution/observation.
Note that all executions, per band, were consecutive and were performed as part of a session.}
\label{table:data_orion}
\bigskip
\end{table*}

\begin{table*}[tbh!]
\centering
\normalsize
\caption{\normalsize Pointings associated with the different Orion-KL mosaic patterns}
\begin{tabular}{lccl}
\hline \noalign {\smallskip}
 Patterns & Band & Nyquist parameter & Pointing \vspace{0.05in} \\
\hline \noalign {\smallskip}
 Hyper-Nyquist & 3 & 4.6 & 1 $\rightarrow$ 15, all pointings \\
 Super-Nyquist & 3 & 3.3 & 2,\,4,\,6,\,8,\,10,\,12,\,14   \\
 Nyquist & 3 & 2.3 & 6,\,8,\,10 \\
 Single pointing & 3 & --- &  8 \\ 
\noalign {\smallskip}
 Hyper-Nyquist & 6 & 4.6 & 1 $\rightarrow$ 45, all pointings\\
 Super-Nyquist & 6 & 3.3 & 2,\,4,\,6,\,8,\,10,\,12,\,14,\,16,\,18,\,20,\,22,\,24,\,26,\,28,\,30,\,32,\,34,\,36,\,38,\,40,\,42,\,44   \\
 Nyquist & 6 & 2.3 &  2,\,4,\,12,\,14,\,22,\,24,\,32,\,34,\,42,\,44 \\
\end{tabular}
\smallskip
\tablecomments{\small Mosaic patterns used in the Band 3 and Band 6 observations. The Nyquist parameter is the ratio of the FWHM of the beam to the shortest distance between two pointings in the pattern.  The pointing numbers correspond to those pictured in Figure \ref{fig:pattern}.
}
\label{table:patterns}
\bigskip
\end{table*}

We calibrate the data using the standard procedures for processing ALMA polarization observations (see, e.g., \citealt{Cortes2016, Nagai2016, Hull2017b, Hull2018a}).  We use J0522--3627 as the polarization calibrator, J0423--0120 as the bandpass calibrator, and J0529--0519 as the gain calibrator. The same calibration sources were used in both the Band 3 and 6 observations.  A bandpass scan was performed only in the first execution of the observing session. 
We derive the bandpass solutions and the flux scaling from both the bandpass calibrator (J0423--0120) and the polarization calibrator (J0522--3627), whereas we correct the amplitude and phase by deriving complex gains from the gain calibrator (J0529--0519).  The polarization calibrator (J0522--3627) was observed every $\sim$\,35 minutes, and had a flux of 8.2\,Jy and $P_\textrm{frac} \approx$\,3.4\% during the Band 3 observations, and a flux of 3.8\,Jy and $P_\textrm{frac} \approx$\,1.7\% during the Band 6 observations.

The mosaicked observations of Orion-KL were set up to allow us to compare different mosaic patterns, which we label according to their different densities of pointings: Hyper-Nyquist (separation between the pointings of $\sim$\,$1/4$ of the FWHM; i.e., over-sampled by a factor of $\sim$\,2 in each dimension relative to a standard Nyquist-sampled mosaic), Super-Nyquist (separation of $\sim$\,$1/3$ of the FWHM; i.e., over-sampled by a factor of $\sim$\,1.5), Nyquist (separation of $\sim$\,$1/2$ of the FWHM; i.e., standard sampling), and a single field (the reference pointing, centered on Source I itself).

The primary beam of each mosaic pointing is assumed (in CASA) to be a 2D Gaussian peaked at the center of the pointing. Figure \ref{fig:pattern} shows the full 3$\times$5 pointing rectangular mosaic pattern of our observations (i.e., the Hyper-Nyquist pattern) at Band 3, and the similar $5\times9$ pattern at Band 6. In the case of Band 3, the field of view of the single pointing includes most of the filament, which allows us to compare the single pointing with the various ALMA mosaics. In contrast, the ALMA Band 6 field of view is too small to allow us to comparison the single pointing and the mosaics.

We image the data using the task \texttt{tclean} from CASA version 5.4.0, using standard imaging parameters and a Briggs visibility weighting of \texttt{robust}\,=\,0.5, resulting in maps with synthesized beams (resolution elements) of 1$\farcs$43\,$\times$\,0$\farcs$95 at Band 3 and 0$\farcs$65\,$\times$\,0$\farcs$48 at Band 6.  
We independently image all four cases (i.e., Hyper-Nyquist, Super-Nyquist, Nyquist, and single field) in order to compare them with one another.  We perform the imaging of all of these cases by specifying the fields that belong to each of the mosaic sampling patterns (see Table \ref{table:patterns}).  For Band 3, all of the mosaicking patterns include the center field.  In the Band 6 data, however, the Nyquist and Super-Nyquist mosaics do not include the center field. Consequently, we re-grid these latter mosaicked images to the Hyper-Nyquist frame using the CASA task \texttt{imregrid} in order to match the coordinates. 

For each of the four patterns, we clean the Stokes $I$, $Q$, and $U$ maps separately. Note that we do not perform self-calibration on any of the images of Orion-KL.  Furthermore, note that spectral lines were not flagged prior to making the continuum images (the default Band 6 spectral setup avoids major lines of interest).  We primary-beam correct the final images using the CASA task \texttt{impbcor} to the 20\% sensitivity level of the Stokes $I$ primary beam model.  We then use these Stokes maps to produce maps of $P$, $P_\textrm{frac}$, and $\chi$.  The mosaicked images of Orion-KL at both Bands 3 and 6 have median levels of $P_\textrm{frac} \approx 6-7\%$ across the maps.  While the signal-to-noise ratio (SNR) of the $P$ images can be >\,30 at the peaks of polarized emission, the median SNR values across the $P$ maps at Bands 3 and 6 are between 7--8.

\section{Results: Orion-KL mosaic observations}
\label{sec:results_orion} 

In order to assess the performance of the ALMA's linear-polarization mosaicking mode in a typical observational scenario, we performed standard (RA,\,DEC) observations of the Orion-KL star-forming region at Bands 3 and 6, and compare the images made using a single pointing versus the three different mosaic patterns (Hyper-Nyquist, Super-Nyquist, and Nyquist).  Table \ref{table:obs_orion} lists the peak fluxes and rms noise levels in the $I$, $Q$, $U$, and $P$ maps for both bands and all sampling patterns. 

\begin{table*}[tbh!]
\centering
\caption{\normalsize Image statistics from Orion-KL observations}
\normalsize
\setlength{\tabcolsep}{0.28em}
\begin{tabular}{lccccccccc}
\hline \noalign {\smallskip}
Pattern & Band & $I_\textrm{peak}$&$I_\textrm{rms}$&$Q_\textrm{peak}$&$Q_\textrm{rms}$&$U_\textrm{peak}$&$U_\textrm{rms}$&$P_\textrm{peak}$&$P_\textrm{rms}$\\
\smallskip
&  & \mjybmvert{} & \mjybmvert{} & \mjybmvert{} & \mjybmvert{} & \mjybmvert{} & \mjybmvert{} & \mjybmvert{} & \mjybmvert{} \vspace{0.05in} \\
\hline \noalign {\smallskip}
Hyper-Nyquist & 3 & 112 & 0.90 & --3.39 & 0.100 & 3.73 & 0.102 & 3.79 & 0.157 \\
Super-Nyquist & 3 & 112 & 0.93  & --3.40 & 0.102 & 3.72 & 0.107 & 3.78 & 0.164 \\ 
Nyquist       & 3 & 113 & 0.96  & --3.47 & 0.118 & 3.75 & 0.120 & 3.81 & 0.180   \\
\smallskip 
Single field  & 3 & 115 & 1.31 & --3.46 & 0.150  & 3.34 & 0.139 & 3.35 & 0.243 \\ 
Hyper-Nyquist & 6 & 415 & 2.08 & 11.3\phantom{0} & 0.522 & 12.3\phantom{00} & 0.495 & 13.5\phantom{00} & 0.485 \\
Super-Nyquist & 6 & 418 & 2.11 & 11.4\phantom{0} & 0.535 & 12.1\phantom{00} & 0.508 & 13.5\phantom{00} &0.519 \\ 
Nyquist       & 6 & 418 & 2.40 & 11.4\phantom{0} & 0.556 & 12.2\phantom{00} & 0.524 & 13.6\phantom{00} & 0.755  \\ 
\end{tabular}
\smallskip
\tablecomments{\small 
The peak intensities and the rms noise levels in the $I$, $Q$, $U$, and $P$ maps derived from the mosaicked and single-field (for Band 3) maps of Orion-KL. 
}
\label{table:obs_orion}
\bigskip
\end{table*}

We first focus our analysis on comparing mosaicked images with the inner $\frac{1}{3}$\,FWHM region of the single pointings, where the effects of residual off-axis errors are minimal.  For both the Band 3 and Band 6 data, we analyze several individual pointings that are common to all three of the mosaics (see Table \ref{table:patterns}) and that show emission in $I$, $Q$, and $U$.  We first produce difference maps (i.e., Hyper-Nyquist minus single pointing, Super-Nyquist minus single pointing, and Nyquist minus single pointing) for $I$, $Q$, and $U$ using the CASA task \texttt{immath}.  We perform our comparison both with and without a cutoff, where the cutoff limits the difference images to the region(s) where the emission in the single pointing images is 5 $\times$ the rms noise level.  We then use the CASA task \texttt{imsubimage} to extract the inner $\frac{1}{3}$\,FWHM regions from the difference maps in order to quantify the impact of overlapping pointings in the mosaics, under the initial assumption that by stitching together pointings to make a mosaic, the emission in the mosaic should be ``polluted'' by the off-axis contributions from neighboring pointings (however, as we see later, we are not actually able to detect the effects of the off-axis pointings in the on-axis data). 

We do not show the distributions of the differences in the Stokes $I$, $Q$, and $U$ maps here because all three difference histograms (for example, the three histograms of the differences in Stokes $Q$ between the chosen single pointing versus the Hyper-Nyquist, Super-Nyquist, and Nyquist mosaics) have similar structure, and have standard deviations that tend to be well below the rms noise level in the single-pointing images.  This suggests that we can consider all of the mosaicked images to be the same when compared with the single pointing.  Note also that the maps are dynamic range limited; in no case have we reached the thermal noise
limit.\footnote{The theoretical sensitivity per pointing in the Band 3 Stokes $I$ maps $\sigma_I$ is estimated to be $\sim$\,30\,\ujybm{} given 7\,min of integration time per field, whereas the estimated noise level in all of the Band 3 mosaic images is $\sim$\,100\,\ujybm{}. We see a similar situation in the Band 6 data, where the theoretical sensitivity $\sigma_I$ is $\sim$\,79\,\ujybm{} given 2.2\,min of integration per field, whereas the estimated noise from the Stokes $I$ maps is $\sim$\,2\,\mjybm{}. The even larger difference between the theoretical versus actual noise in the Band 6 maps is not surprising given the higher angular resolution of the Band 6 data (and thus limited ability to recover extended structure, and larger resultant imaging artifacts) relative to the Band 3 data.} 
These findings allow us to draw two main conclusions: first, the errors in the inner $\frac{1}{3}$\,FWHM of the $Q$ and $U$ (and also $I$) maps are primarily caused by imaging artifacts stemming from the inability of our ALMA observations to recover emission at large spatial scales in Orion-KL, rather than by residual off-axis polarization errors; if the latter were dominant, we should expect to see a change in the width of the $Q$ and $U$ difference histograms with increased mosaic packing, since increased packing should reduce the contribution from off-axis errors.  Second, while packing the pointings in the mosaics more closely should in theory improve the accuracy of the polarization images, we are not able to detect these incremental improvements among the Nyquist, Super-Nyquist, and Hyper-Nyquist mosaics.  Thus, based on our dynamic-range-limited Orion-KL images (and for other sources with similarly complex, multi-scale structure), we cannot recommend using a mosaic packing that is tighter than the standard Nyquist pattern.

When analyzing the Orion-KL results, we choose not to analyze the fractional polarization errors, as the fractional polarization is equal to $\sqrt{Q^2 + U^2} / I$; with respect to Stokes $I$, Stokes $Q$ and $U$ have different physical origins and thus different spatial structure (including different distributions of power as a function of spatial scale) when the source is resolved \citep[for example, see the polarized intensity $P$ versus Stokes $I$ emission in high-resolution ALMA polarization maps in, e.g.,][]{Maury2018, Cortes2019, Hull2020a, LeGouellec2019}.  Finally, the Stokes $Q$ and $U$ maps have very different dynamic range limitations compared with the Stokes $I$ maps.  Consequently, characterization of polarization fraction errors is better performed with point sources, as we have done in the 11$\times$11 observations of 3C~279 (see Section \ref{sec:11x11_errors}).  

We thus choose only to use the polarization position angle $\chi$ to characterize the effects of residual off-axis errors in the Orion-KL mosaics.  Below we analyze $\chi$ only for the lower-resolution Band 3 data, because of the simplicity afforded by the larger field of view relative to Band 6, and because the large-scale structure in the Band 3 data is significantly less spatially filtered than in the higher resolution Band 6 data, resulting in lower-level imaging artifacts.\footnote{The Band 6 mosaics have strong negative and positive ``bowls'' and ``ridges'' at the edges of the image, parallel to the main axis of the star-forming filament.  These artifacts are the result of the inability of our ALMA observations to recover large-scale structure (i.e., short $uv$-spacings) toward the Orion-KL filament, which is highly complex, and has significant emission at spatial scales larger than those recoverable by ALMA.  These artifacts would be considerably attenuated if we were to fill in the missing $uv$-spacings by combining the ALMA observations with total-power (single-dish) observations, but we do not attempt this here.  For a description of this phenomenon, see Section 11.5.2 of \citet{TMS}.}

\subsection{Comparison of $\chi$ in the mosaics versus the inner region ($r~<~\frac{1}{3}$\,FWHM) and outer ``donut'' region ($\frac{1}{3}$\,FWHM~$<~r~<$~FWHM) of single pointings}
\label{sec:inner_1/3_vs_donut}

\begin{figure}[tbh!]
\centering
\includegraphics[scale=0.41, clip, trim=0cm 0cm 0.5cm 0.0cm]{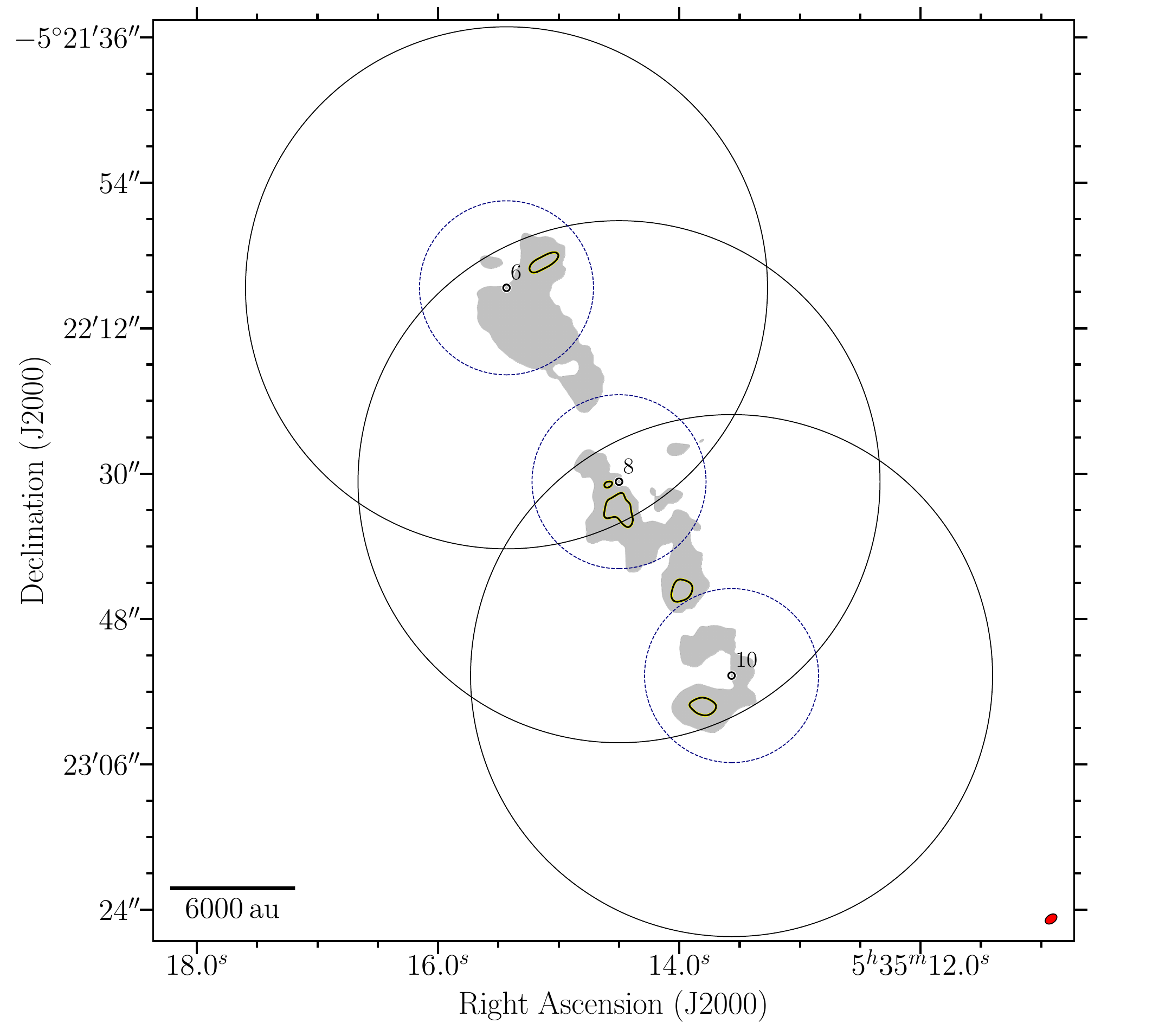}
\caption{ \small 
Setup of the test to compare the polarization position angles in the three mosaics with those in the ``donut''-shaped region outside of the $\frac{1}{3}$\,FWHM in the three single pointings comprising the Band 3 Nyquist mosaic.  The gray shaded area indicates the region where polarized emission is detected in the hyper-Nyquist mosaic map of Orion KL, and is masked below 5\,$\sigma_P$, where the rms noise in the $P$ map $\sigma_P = 0.15$\,\mjybm{}.  Thick black contours indicate the regions where the $P$ map has SNR\,>\,15.  The concentric circles show the FWHM (large circles) and the $\frac{1}{3}$\,FWHM (small circles) for the three different single pointings.  The synthesized beam is shown as an ellipse in the lower-right corner.
}
\bigskip
\label{fig:donut}
\end{figure}

We compute difference maps of the polarization position angle $\chi$ in the Hyper-Nyquist mosaic versus single-pointing images in the same manner as described above.  The three single pointings we use are those in the Band 3 Nyquist mosaic; the setup of the test can be seen in Figure \ref{fig:donut}.  We use a conservative (5\,$\sigma_P$) cutoff in the $\chi$ maps in order to avoid considering any spurious points at the edge of the maps in our statistics. 

\begin{figure*}[tbh!]
\centering
\includegraphics[scale=0.27, clip, trim=0.5cm 0cm 1cm 1.5cm]{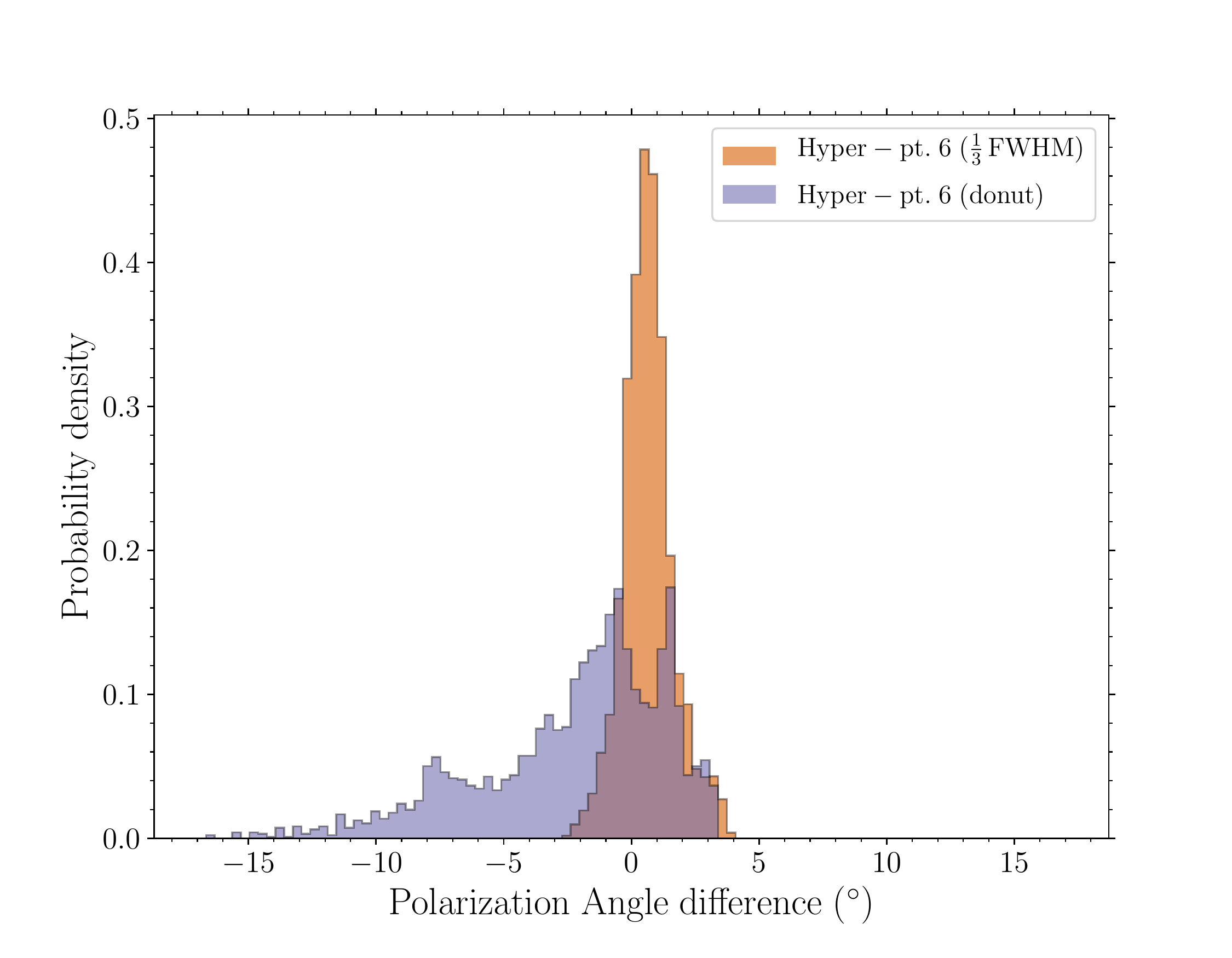}
\includegraphics[scale=0.27, clip, trim=0.5cm 0cm 1cm 1.5cm]{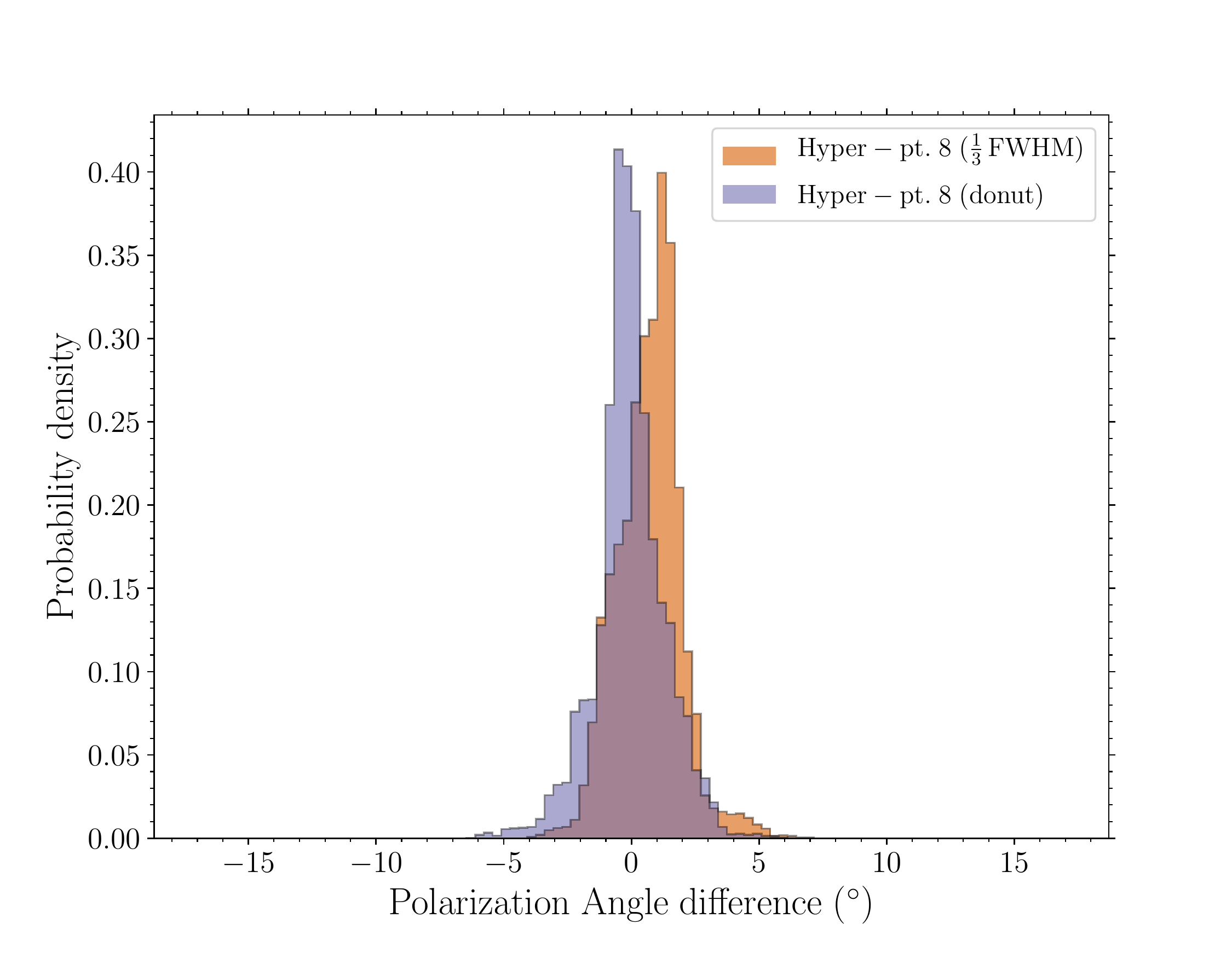}
\includegraphics[scale=0.27, clip, trim=0.5cm 0cm 1cm 1.5cm]{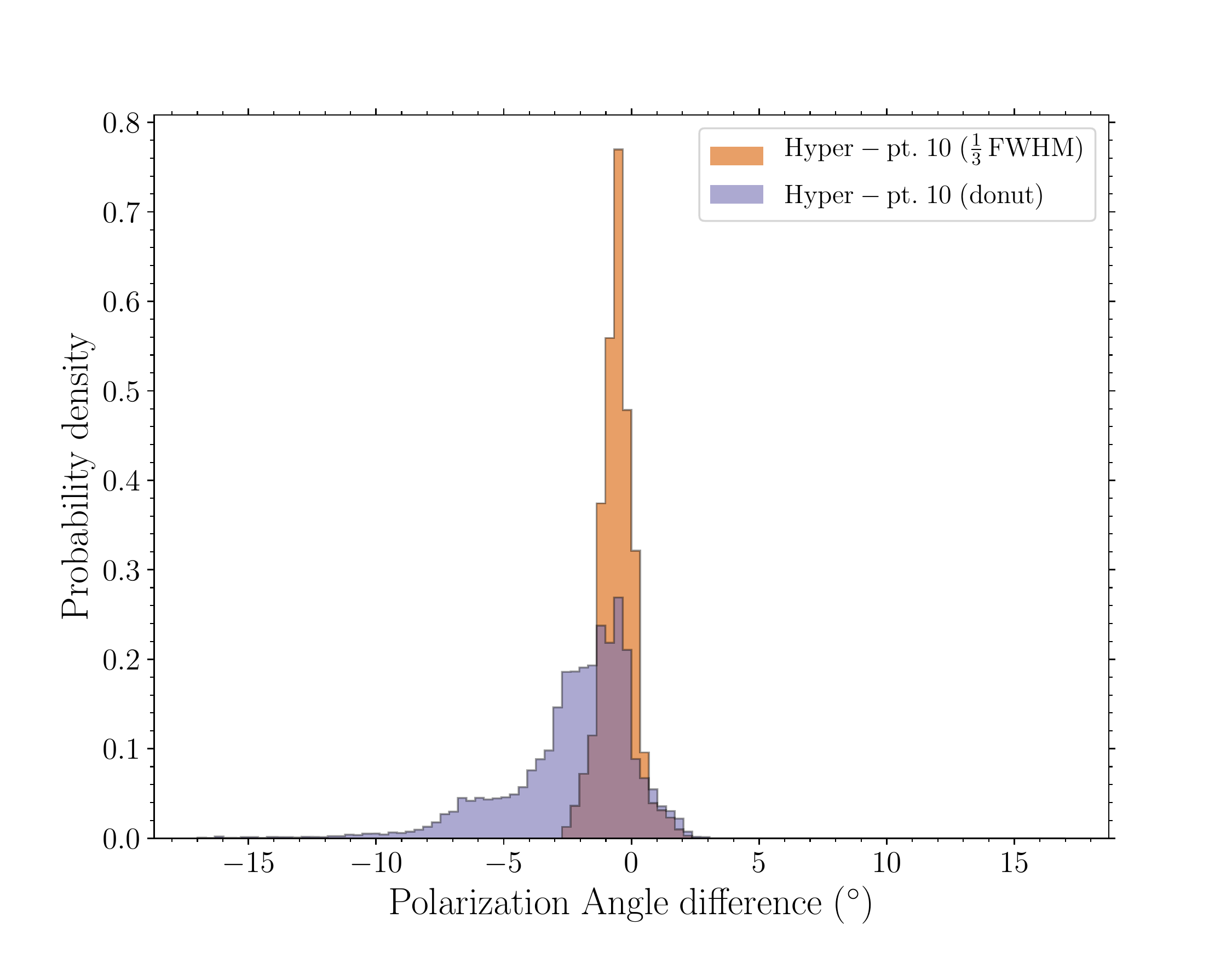}
\caption{
Difference maps between the polarization position angle $\chi$ from three individual pointings in the Band 3 observations versus the same regions in the Hyper-Nyquist mosaic.  The histograms show the differences in the inner $\frac{1}{3}$\,FWHM as well as in the outer ``donut'' region ($\frac{1}{3}$\,FWHM $< r <$ FWHM).}
\bigskip
\label{fig:diffMapsPOLA}
\end{figure*}

The histograms of the $\chi$ differences within the inner $\frac{1}{3}$\,FWHM of the single pointings can be seen in Figure \ref{fig:diffMapsPOLA}.  They all have standard deviations of <\,1.3$\degree$, revealing that at the $\sim$\,1$\degree$ level, the regions of the mosaic corresponding to the inner $\frac{1}{3}$\,FWHM region of a given pointing are not, in fact, ``polluted'' by the off-axis regions of the neighboring pointings, as we had initially assumed.  

In Figure \ref{fig:diffMapsPOLA} we also show the histograms of the $\chi$ differences in the outer ``donut'' regions ($\frac{1}{3}$\,FWHM~$<~r~<$~FWHM).  It is clear that in the two outer pointings (point 6 and point 10), the histograms of the differences in $\chi$ in the ``donut'' regions are significantly wider than in the inner $\frac{1}{3}$\,FWHM regions, having standard deviations of 3.6$\degree$ (point 6) and 2.7$\degree$ (point 10).  This suggests that we can detect the effects of residual off-axis errors in these single-pointing maps, consistent with what was seen in the 11$\times$11 maps.

Mosaicking reduces these off-axis errors as a result of overlapping multiple pointings.  A visually simple demonstration of this error-canceling effect can be seen by looking at Figure \ref{fig:donut}.  In all three pointings, the median differences in $\chi$ in the high-SNR patches of polarized emission (see the thick contours in Figure \ref{fig:donut}, which inducate regions where $P$ has an SNR\,>\,15, corresponding to a statistical uncertainty of $\sim$\,2$\degree$)---when observed by the Hyper-Nyquist mosaic versus when observed \textit{on-axis} in the single pointings---are always <\,1$\degree$.  However, for example, when the patch of emission near the center of point 6 is observed near the FWHM of point 8 (i.e., when the emission is observed off-axis), the median difference with respect to the Hyper-Nyquist mosaic is $\sim$\,3$\degree$, as expected from the Band 3 11$\times$11 tests (see Figure \ref{fig:errors_B3-3}).  This same effect is also seen when the other patches of high-SNR emission near the centers of points 8 and 10 are observed off-axis: the median position angle difference with respect to the Hyper-Nyquist mosaic when the emission is observed off-axis in the single pointings is always larger than the difference when the same emission is observed on-axis.  These simple tests demonstrate again that mosaicking the image reduces the off-axis polarization angle errors.  

Note that, in an effort to connect these Orion-KL error results with the errors we derived from the 11$\times$11 blazar observations (see Section \ref{sec:11x11_errors}), we perform a statistical analysis of the 11$\times$11 data.  This analysis, which is an attempt to estimate the errors in a mosaic from the 11$\times$11 data, can be found in Appendix \ref{app:error_11x11}.

\subsection{Comparison with CARMA}
\label{sec:carma}

\begin{figure*}[hbt!]
\centering
\includegraphics[scale=0.49, clip, trim=0.2cm 1.0cm 3cm 2.5cm]{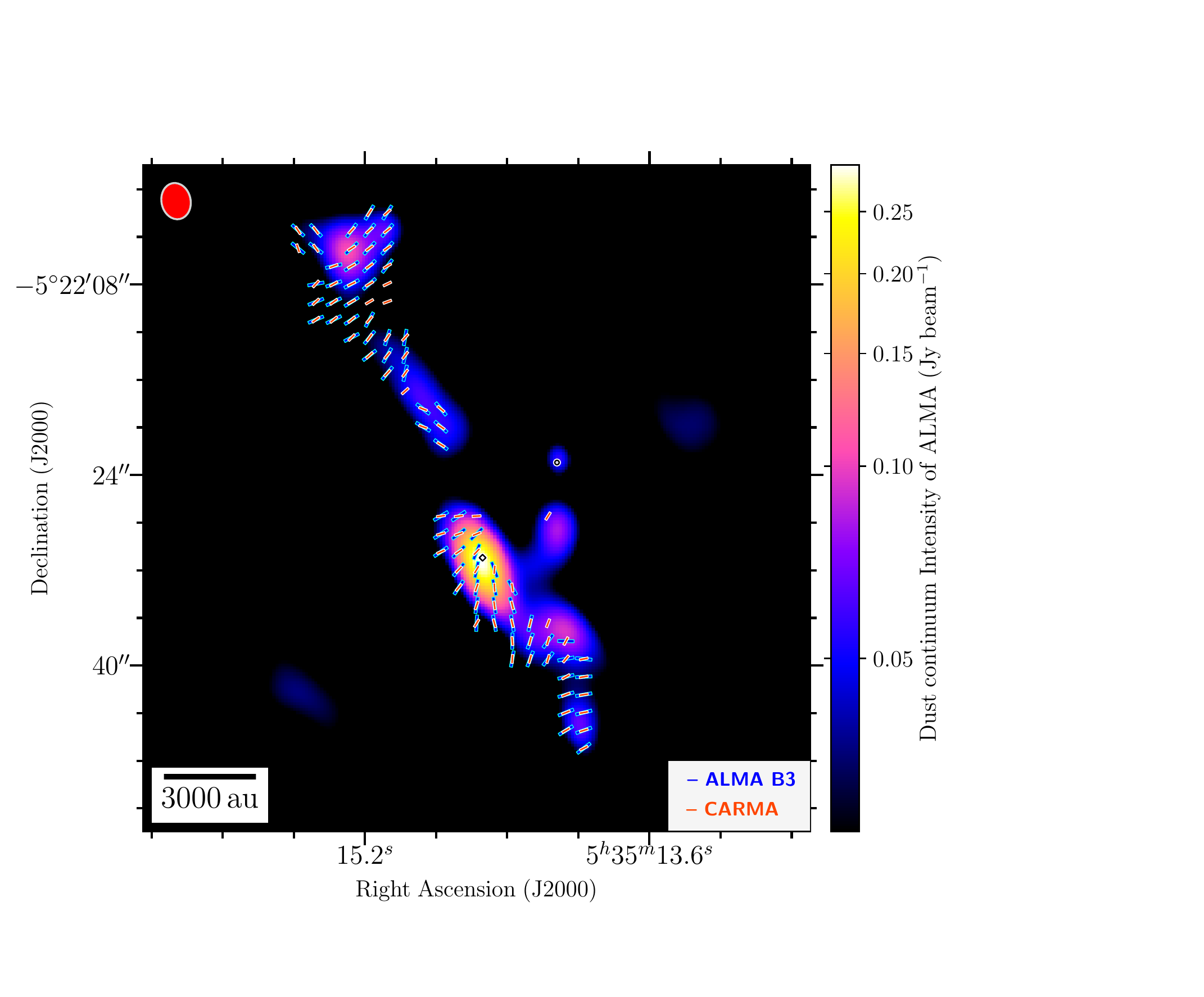}
\includegraphics[scale=0.49, clip, trim=0.2cm 1.0cm 4cm 2.5cm]{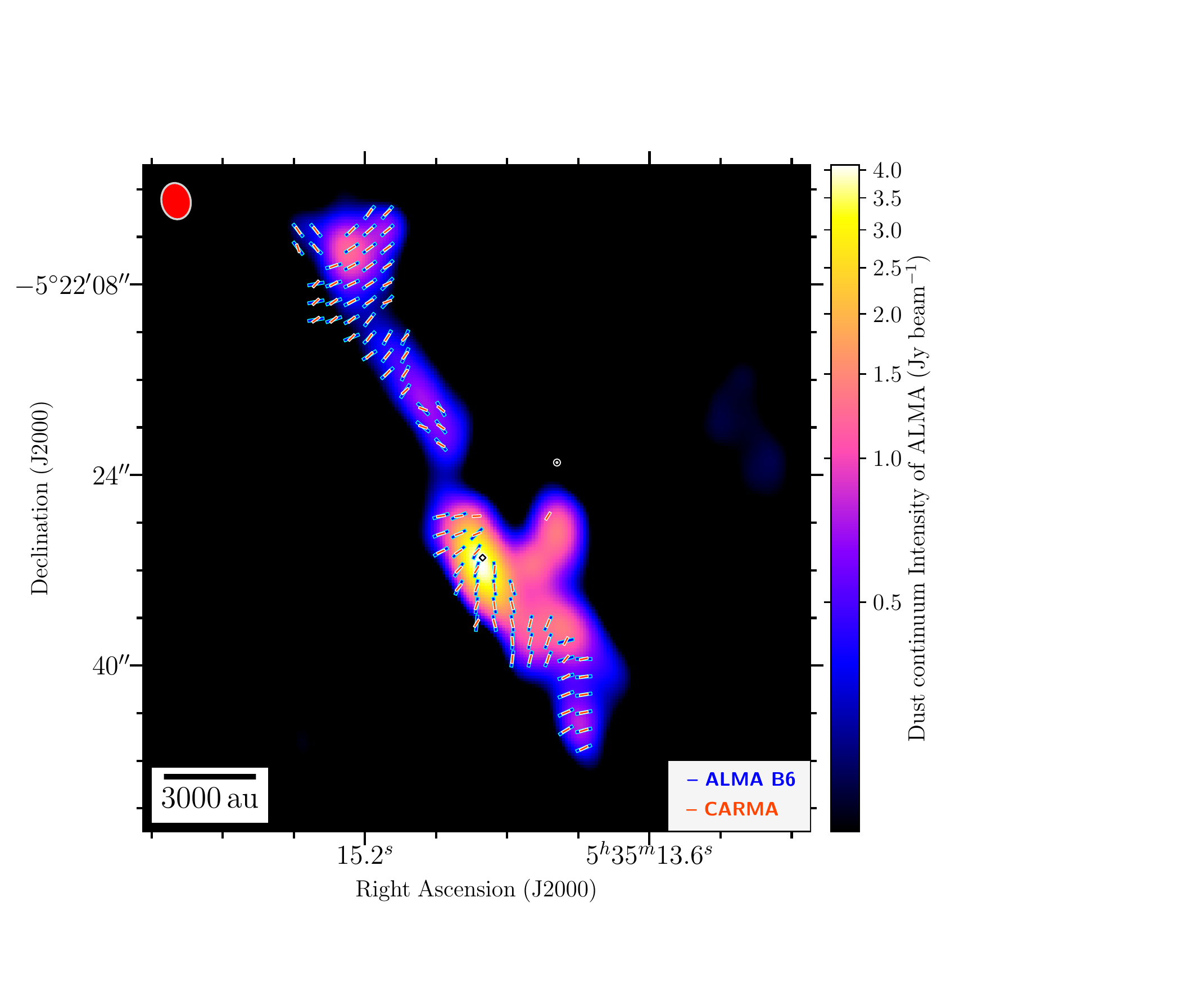}
\caption{\small 
Comparison of ALMA and CARMA observations of the inferred magnetic field (rotated by 90$\degree$ with respect to the observed polarization) toward Orion-KL using the Nyquist-sampled ALMA data (blue line segments) and CARMA data (orange line segments).  Both maps show the same 1.3\,mm CARMA data; the left- and right-hand maps show the Band 3 (3\,mm) and Band 6 (1.3\,mm) ALMA data, respectively.  We produce both ALMA maps using the same $uv$-coverage as the CARMA observations, and then smooth them to the CARMA beam size of 3$\farcs$07$\,\times\,$2$\farcs$45 (shown as ellipses in the upper-left corners of both plots).  The color scales show the Stokes $I$ thermal dust emission from ALMA, which we plot beginning at 3\,$\sigma_I$, where $\sigma_I$ is the rms noise in the smoothed ALMA maps.  $\sigma_I = 8.9$\,\mjybm at Band 3 and 55\,\mjybm at Band 6.  The CARMA polarization map is the one provided in the online version of \citet[][see their Figure 19]{Hull2014}; the map was already masked, and thus we perform no additional cuts.
We plot the ALMA polarization detections only in locations with a corresponding CARMA detection. The diamond and circle indicate the positions of Source I and the Becklin-Neugebauer Object \citep[BN;][]{Becklin1967}, respectively, plotted at the positions reported in \citet{Rodriguez2017}.  The distance to the Orion Nebula Cluster is $388 \pm 5$\,pc \citep{Kounkel2017}.
}
\label{fig:CARMA_mag}
\bigskip
\bigskip
\end{figure*}

Another mosaicked polarization observation of Orion-KL was performed by \citet{Hull2014} using the CARMA interferometer's full-polarization system \citep{Hull2015b}.  Here we compare those archival data from an independent instrument with our ALMA results in order to characterize the performance of ALMA linear-polarization mosaics.  The 1.3\,mm CARMA data were taken in 2013, have a resolution of 3$\farcs$07$\,\times\,$2$\farcs$45, and comprise a seven-pointing mosaic with a field of view similar to that of our ALMA observations. In order to perform the most accurate possible comparison between the CARMA and ALMA datasets, we match the $uv$-range (using the \texttt{uvrange} keyword in the CASA task \texttt{clean}) and the synthesized-beam size (using the CASA task \texttt{imsmooth}) of the Band 3 and Band 6 ALMA data to those of CARMA. We also re-grid the ALMA images to match the coordinates of the CARMA maps.  We use the standard Nyquist mosaic pattern from our ALMA data for the comparison with the CARMA observations, which are also Nyquist sampled. 

See Figure \ref{fig:CARMA_mag} for an overlay of the inferred magnetic field from the CARMA data with both the Band 3 and Band 6 ALMA data.  We can see that the magnetic field orientation in both the Band 3 and the Band 6 ALMA maps matches the CARMA observations remarkably well, especially in the Northern Ridge.  This consistency of inferred magnetic field orientation as a function of frequency is expected in optically thin material (which, we should note, might not be the case toward the very central region near Source I), under the standard assumption that dust-grain alignment via radiative alignment torques \citep[RATs;][]{Lazarian2007} is the cause of the dust polarization.

\begin{figure*}[hbt!]
\centering
\includegraphics[scale=0.48, clip, trim=0.2cm 1.5cm 3cm 2cm]{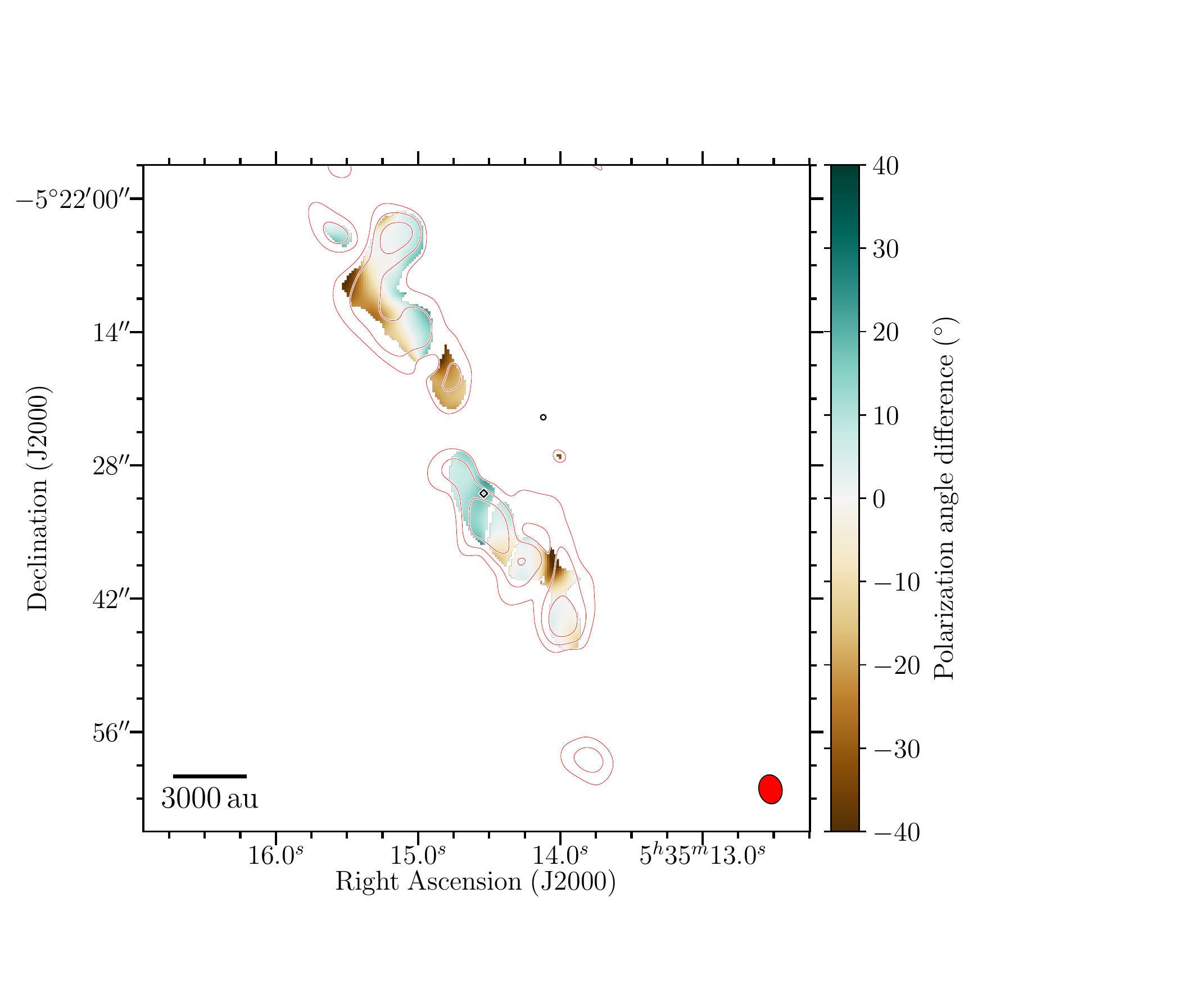}
\includegraphics[scale=0.43, clip, trim=0.5cm 0.5cm 2cm 2cm]{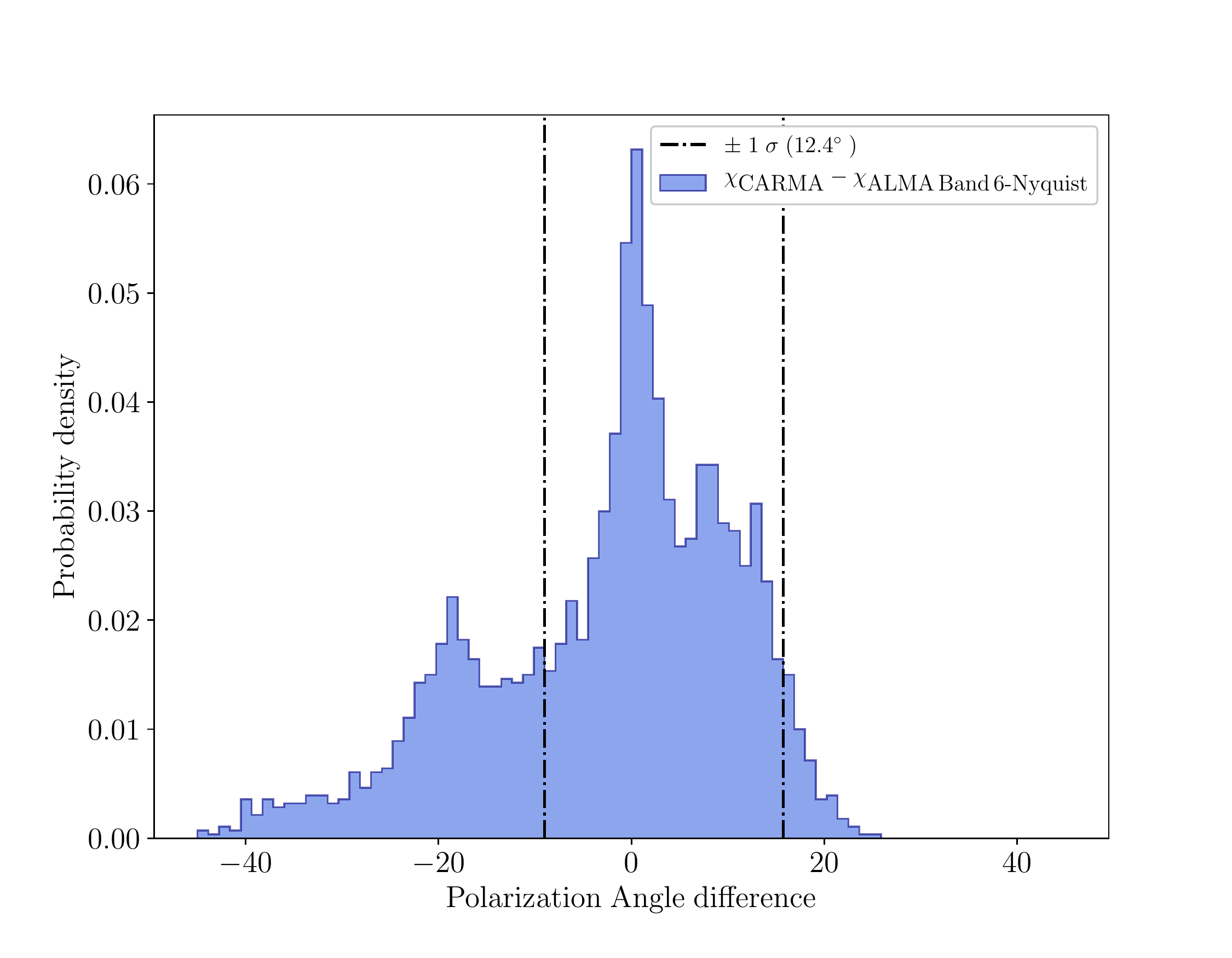}
\caption{\small 
\textit{Left:} Map of the differences in the polarization position angle $\chi$ between the Nyquist-sampled ALMA Band 6 image of Orion-KL and the 1.3\,mm CARMA map from \citet{Hull2014}. The red contours show the polarized intensity $P$ in the ALMA data, which we plot at 3,\,5,\,9\,$\times$\,$\sigma_P$, where $\sigma_P = 3.9$\,\mjybm{} in the smoothed ALMA map, as derived from the Python \texttt{astropy.stats.mad\_std} method.  Note that as in Figure \ref{fig:CARMA_mag}, we normalize $\sigma_P$ by the primary-beam-correction map.  We plot the diamond, circle, and ellipse (synthesized beam) as in Figure \ref{fig:CARMA_mag}.
\textit{Right:} histogram of the differences in $\chi$ in the CARMA versus ALMA maps.  We only include pixels whose polarized intensity is $\geq$\,5\,$\times$\,$\sigma_P$ of the ALMA data. The map from \citet{Hull2014} was already masked based on the SNR of its polarized intensity. The two black, dot-dashed lines encompass 68\% of the distribution; the width of the distribution, defined as half the distance between the two lines, is $\sim$\,12$\degree$.
}
\label{fig:CARMA_diff}
\bigskip
\bigskip
\end{figure*}

In Figure \ref{fig:CARMA_diff} we show both a map and a histogram of the differences in the polarization position angle $\chi$ in the ALMA versus the CARMA data.  The distribution of differences peaks around 0$\degree$, with a width (defined as the minimum window necessary to encompass 68\% of the data) of $\pm$\,12$\degree$.  The differences in $\chi$ can be as large as $\pm$\,40$\degree$ in some locations, and the distribution is somewhat skewed to negative differences, but there is no obvious systematic trend in angle differences in the CARMA versus ALMA maps.  The ALMA polarization position angles have a systematic error of <\,1$\degree$ on-axis \citep{Nagai2016}.  The CARMA polarization position angle measurements have a systematic on-axis error of $\pm$\,3$\degree$, which is the result of systematic errors in the $XY$-phase correction (including variations in the absolute position angle of the wire grids used to derive the $XY$-phase passbands; \citealt{Hull2015b}).

\section{Conclusions}
\label{sec:conc}

We characterize the wide-field errors in interferometric observations performed using the ALMA array of 12\,m antennas.  We first measure the errors across the field of view of a single-pointing image by observing the highly linearly polarized blazar 3C~279 in 11$\times$11 grids out to the FWHM of primary beam in the (Azimuth, Elevation) frame.  Next, in order to characterize ALMA's polarization mosaic performance during a standard-mode observation in the (RA, DEC) frame, we performed mosaicked linear-polarization observations at Bands 3 and 6 of the Orion-KL star-forming region using several different mosaic patterns.  

These are the main conclusions we draw from the 11$\times$11 observations of the highly linearly polarized blazar 3C~279:

\begin{outline}

\1 [1.] After on-axis calibration of all off-axis pointings, we find that:

\2 [(a)] The systematic errors in polarization fraction ($P_\textrm{frac,err} = P_\textrm{frac,on} - P_\textrm{frac,off}$) for all observed bands (3, 5, 6, 7) are $\lesssim$\,0.001 ($\lesssim$\,0.1\% of Stokes $I$) near the beam center, and increase to $\sim$\,0.003--0.005 ($\sim$\,0.3--0.5\% of Stokes $I$) near the FWHM of the primary beam.  

\2 [(b)] The systematic errors in the polarization position angle ($\chi_\textrm{err} = \chi_\textrm{on} - \chi_\textrm{off}$) are $\lesssim$\,1$\degree$ near the beam center, and $\sim$\,1--5$\degree$ near the FWHM.

\1 [2.] In all bands, we see the expected double-lobed ``beam squint'' patterns in the Stokes $V$ maps at the 1--2\% level of Stokes $I$.

\1 [3.] The off-axis errors in the $\chi$ and $P_\textrm{frac}$ maps arise from differences in the shapes of the $Q$ and $U$ beams.  Also known as ``beam squash,'' this effect causes cloverleaf patterns of positive and negative lobes in the $Q$ and $U$ error maps.  The quadrupolar squash error patterns are approximately twice the linear extent of the double-lobed beam squint patterns, and thus our 11$\times$11 observations only sample the inner part of the squash error patterns.

\end{outline}

Next, we analyze linear-polarization mosaics of the Orion-KL star-forming region observed using several patterns: a single-pointing observation, a standard Nyquist-sampled mosaic, and two more densely packed (``Super-Nyquist'' and ``Hyper-Nyquist'') mosaics.  While we conclude that mosaicking improves the accuracy of the polarization images relative to the single pointing images, we are not able to detect incremental improvements in the accuracy among the three mosaics.  Thus, based on our Orion-KL images (and for other sources with similarly complex, multi-scale structure), we cannot recommend using a mosaic packing that is tighter than the standard Nyquist pattern.  Finally, we compare the ALMA results with archival CARMA data.  Our conclusions from these efforts are as follows:

\begin{outline}

\1 [4.] We compare the Band 3 and Band 6 mosaicked images of Orion-KL with the inner $\frac{1}{3}$\,FWHM region of several single pointings.  We find that the errors in the inner $\frac{1}{3}$\,FWHM of the $Q$ and $U$ (and also $I$) maps are primarily caused by imaging artifacts stemming from the inability of our ALMA observations to recover emission at large spatial scales in Orion-KL, rather than by residual off-axis polarization errors.  Furthermore, the different spatial structure of the $I$ versus $Q$ and $U$ emission makes it difficult to perform a meaningful analysis of the differences in maps of $P_\textrm{frac}$.  Characterization of errors in $P_\textrm{frac}$ is thus better performed with point sources, as we have done in the 11$\times$11 observations of 3C~279.  

\1 [5.] The polarization position angle $\chi$ is a better quantity for characterizing off-axis errors, as it is independent of $I$.  The difference maps of $\chi$ in the (Band 3 only) Hyper-Nyquist mosaic versus the inner $\frac{1}{3}$\,FWHM regions of three Band 3 single pointings show that the differences are $\sim$\,1$\degree$, indicating that, at the $\sim$\,1$\degree$ level, the inner regions of a given pointing are not ``polluted'' by the errors in the off-axis regions of neighboring pointings.

\1 [6.] Next we compare the (Band 3 only) Hyper-Nyquist mosaic with the outer ($\frac{1}{3}$\,FWHM < $r$ < FWHM) ``donut'' region of the same three Band 3 single pointings.  We find that, in two of the three pointings, the distributions of the differences between $\chi$ in the outer ``donut'' regions of the single pointings versus the mosaic are significantly wider than the differences in the inner $\frac{1}{3}$\,FWHM of the single pointings versus the mosaic.  This suggests that we can detect the effects of off-axis errors in the single-pointing maps, consistent with what was seen in the 11$\times$11 tests.

\1 [7.] We perform a simple test using the Band 3 Orion-KL data where we analyze several patches of high-SNR polarized emission.  In all three pointings, the median differences in $\chi$ in the high-SNR patches of polarized emission---when observed by the Hyper-Nyquist mosaic versus when observed \textit{on-axis} in the single pointings---are always <\,1$\degree$.  However, when the same patches of emission are viewed off-axis in adjacent single pointings, the median position angle differences with respect to the Hyper-Nyquist mosaic are always larger (with typical median values of 2--4$\degree$) than when the same emission is observed on-axis.  These larger differences are consistent with the Band 3 11$\times$11 tests.  These tests demonstrate again that mosaicking reduces off-axis errors.   

\1 [8.] Finally, we perform a comparison of both the 3\,mm and 1.3\,mm ALMA mosaics with a 1.3\,mm CARMA mosaic from \citet{Hull2014}.  After matching the $uv$-coverage of the ALMA and CARMA data and then smoothing the ALMA data to the coarser CARMA resolution, we find good consistency in the polarization position angles in both bands, especially in the Northern Ridge region of Orion-KL.  The distribution of differences in $\chi$ in the CARMA versus the ALMA maps has a width of $\pm$\,12$\degree$.

\end{outline}

Our results show that, due to the large angular extent of the beam-squash pattern of the ALMA 12\,m antennas, the errors introduced by this wide-field effect are modest.  We show that mosaicking reduces the magnitudes of these errors because multiple pointings are overlapped, and because the mosaic-imaging algorithm puts a higher weight on emission that is on-axis versus emission that is far from the pointing center.  However, ultimately we would like to be able to calibrate out these effects, thus yielding accurate wide-field polarization images even in single-pointing observations.  Efforts to implement full-polarization voltage-pattern corrections (i.e., primary beam models) at ALMA are underway, and will allow us to achieve this goal in the future.

\acknowledgments
The authors thank the anonymous referee for the positive commentary and helpful suggestions.  
The authors thank Anita Richards for the useful commentary on the manuscript, and for the help preparing the data release.
C.L.H.H. acknowledges Sanjay Bhatnagar and Preshanth Jagannathan for the thoughtful discussion.
C.L.H.H., P.C.C., S.K., E.B.F., and C.L.B. acknowledge the support of the ALMA Development Program, which funded the Cycle 5 ALMA Development Study, ``Full-Mueller Mosaic Imaging with ALMA'' (PI: S. Bhatnagar).
At the inception of this project, C.L.H.H. was a Jansky Fellow of the National Radio Astronomy Observatory, which is a facility of the National Science Foundation operated under cooperative agreement by Associated Universities, Inc.  
C.L.H.H. acknowledges the support of both the NAOJ Fellowship as well as JSPS KAKENHI grants 18K13586 and 20K14527.
V.J.M.L.G. acknowledges the support of the ESO Studentship Program.
This paper makes use of the following ALMA data: ADS/JAO.ALMA\#2011.0.00007.E and ADS/JAO.ALMA\#2011.0.00008.E (Orion-KL mosaics), and ADS/JAO.ALMA\#2011.0.00009.E (11$\times$11 data).  ALMA is a partnership of ESO (representing its member states), NSF (USA) and NINS (Japan), together with NRC (Canada), MOST and ASIAA (Taiwan), and KASI (Republic of Korea), in cooperation with the Republic of Chile. The Joint ALMA Observatory is operated by ESO, AUI/NRAO and NAOJ.

\smallskip 
\textit{Facilities:} ALMA.

\smallskip
\textit{Software:} APLpy, an open-source plotting package for Python hosted at \url{http://aplpy.github.com} \citep{Robitaille2012}.  CASA \citep{McMullin2007}.  Astropy \citep{Astropy2018}.  

All of the reduction and imaging scripts used to produce the results from this paper are publicly available.  They can be found, along with the raw data, by visiting the ALMA Science Archive and searching for the ALMA project codes listed in the acknowledgements section.

\bibliography{ms}
\bibliographystyle{apj}

\appendix
\addcontentsline{toc}{section}{Appendix}
\renewcommand{\thesubsection}{\Alph{subsection}}


\subsection{\ref{app:data}. Off-axis Stokes $I,Q,U,V$ data using the highly linearly polarized blazar 3C~279}
\label{app:data}

\begin{figure} [H]
\centering
\includegraphics[scale=0.58, clip, trim=0.9cm 0cm 0cm 0.9cm]{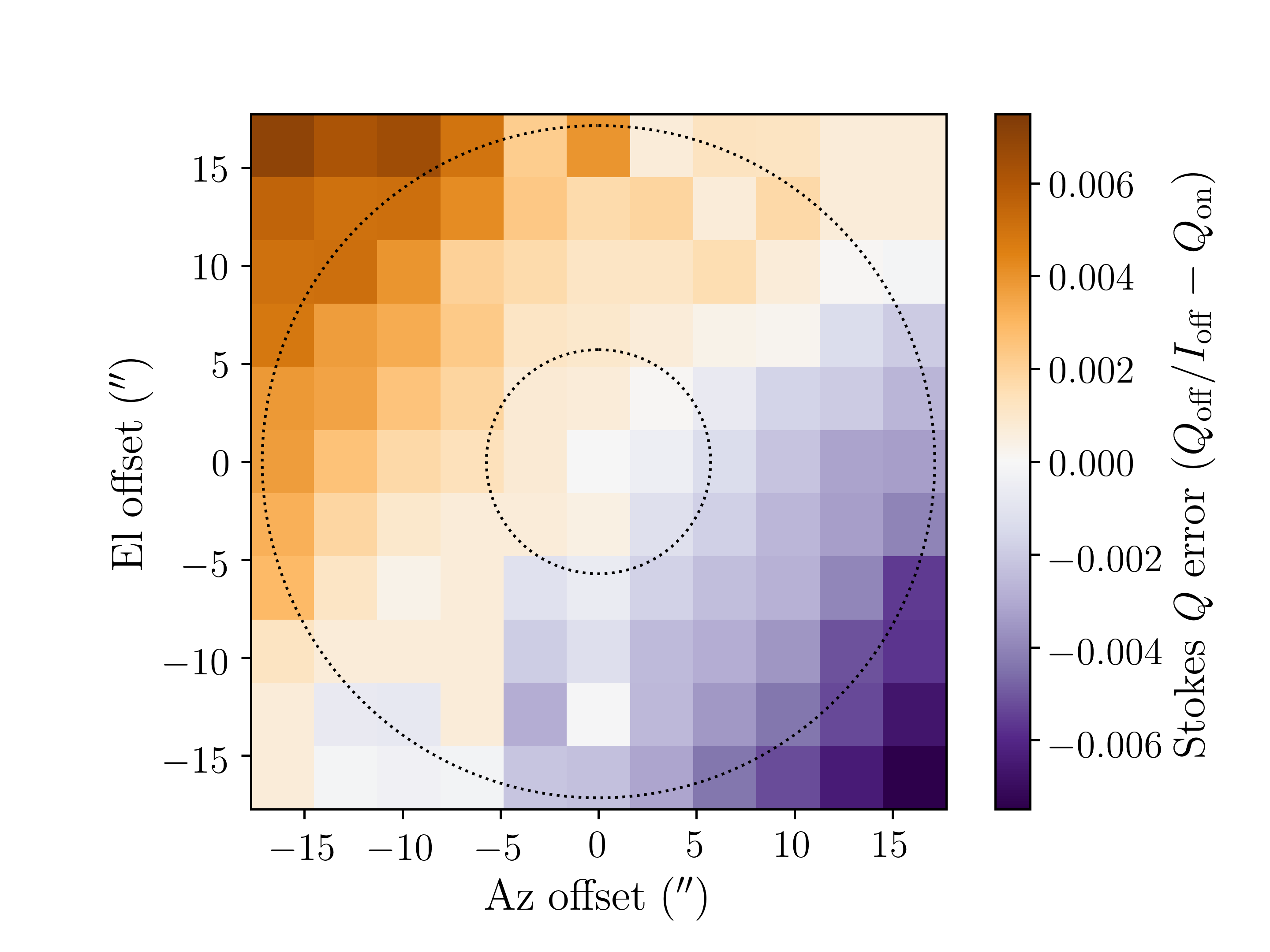}
\includegraphics[scale=0.58, clip, trim=0.9cm 0cm 0cm 0.9cm]{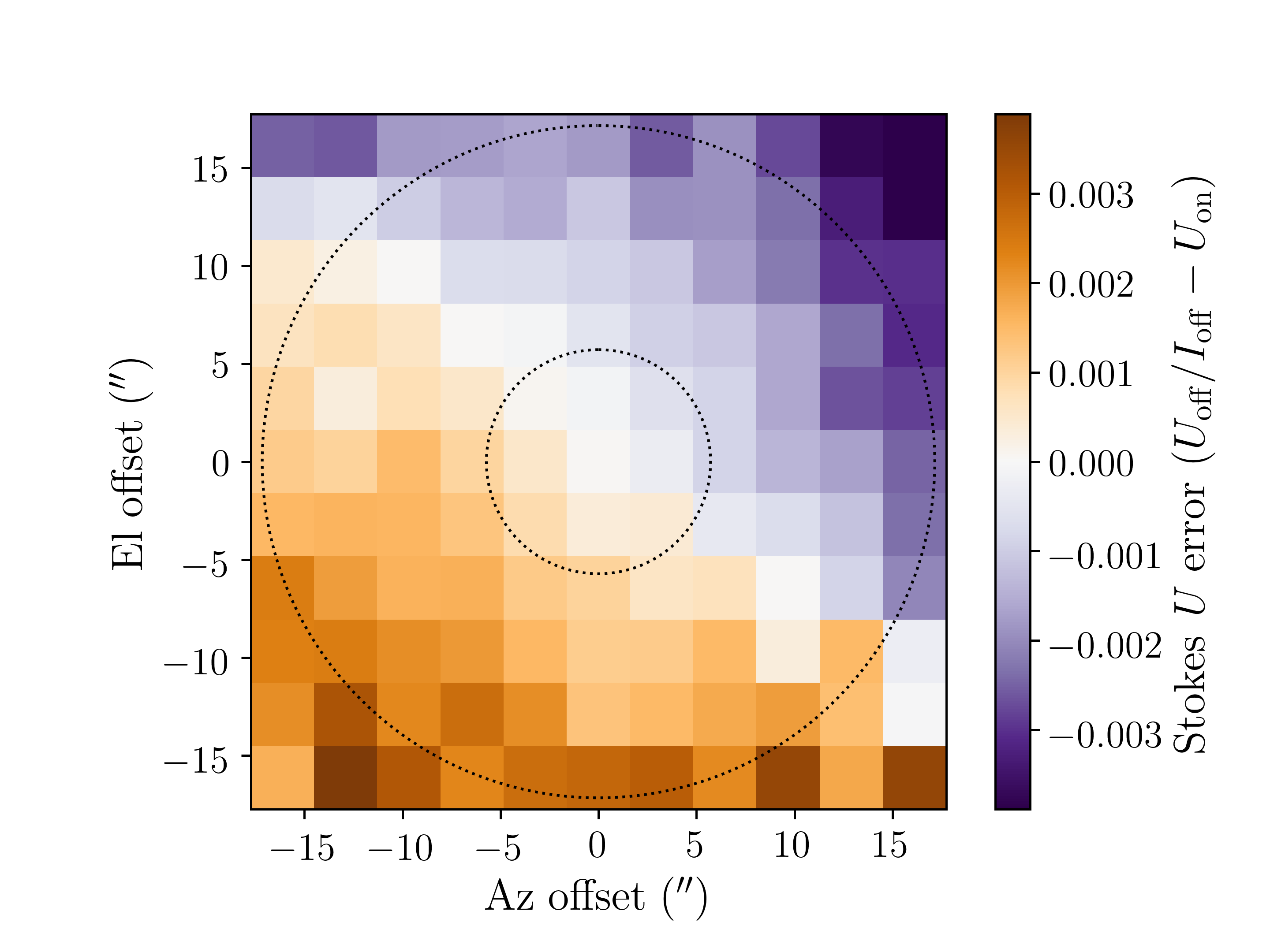} \\
\includegraphics[scale=0.58, clip, trim=0.9cm 0cm 0cm 0.9cm]{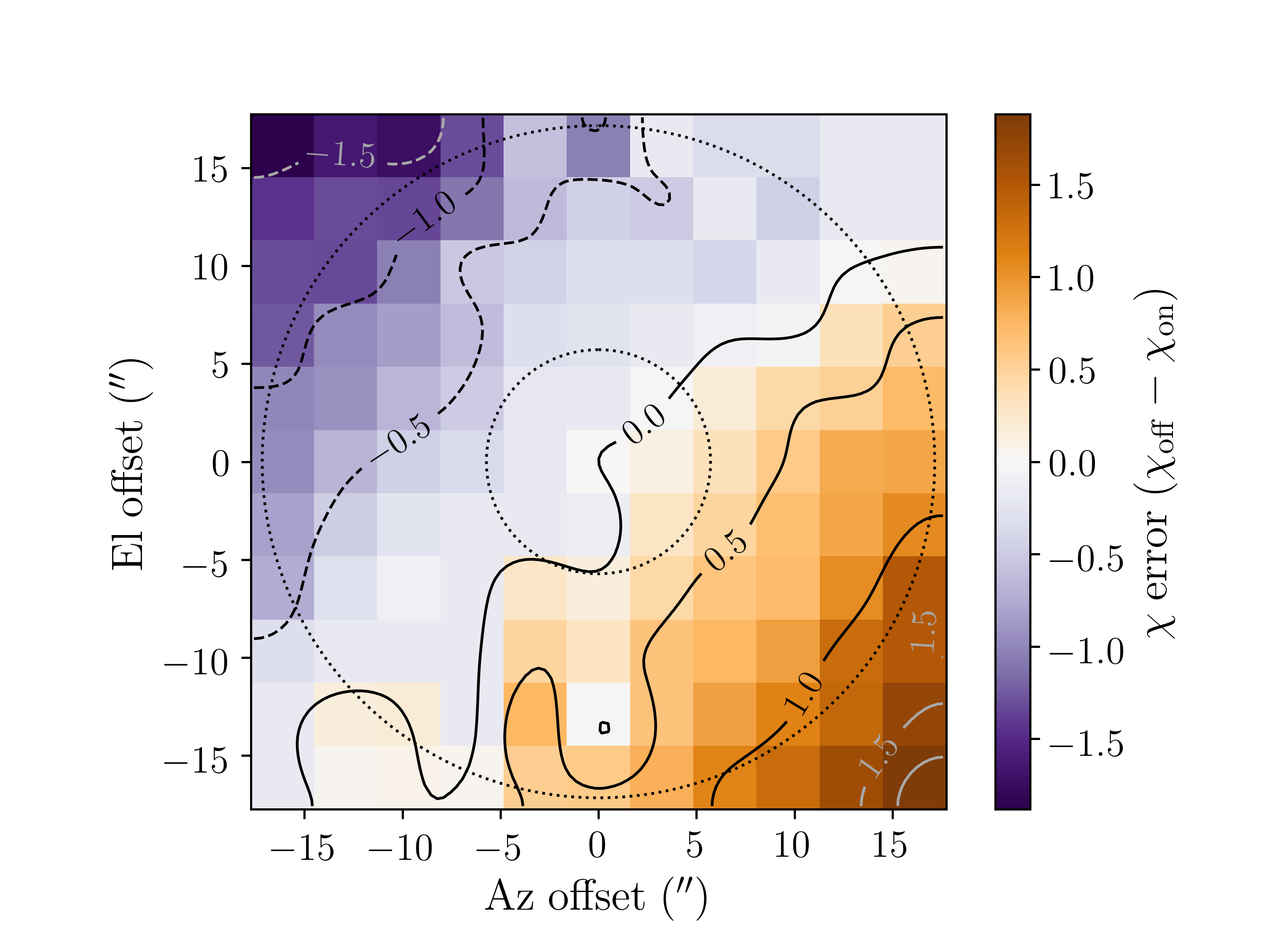}
\includegraphics[scale=0.58, clip, trim=0.9cm 0cm 0cm 0.9cm]{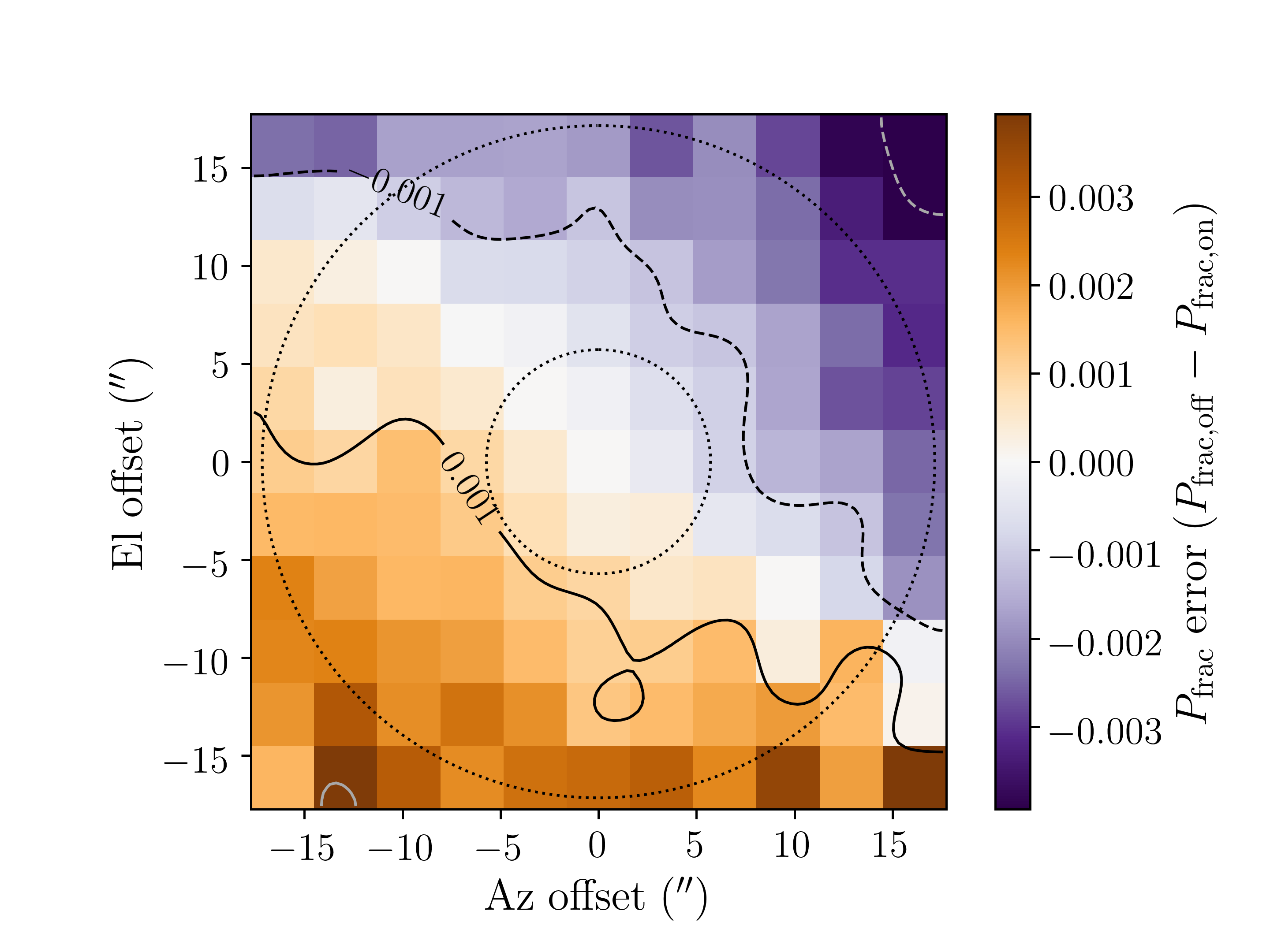}
\caption{\small Same as Figure \ref{fig:errors_B3-3}, for Band 5.}
\label{fig:errors_B5-1}
\end{figure}

\begin{figure*} [hbt!]
\centering
\includegraphics[scale=0.58, clip, trim=0.9cm 0cm 0cm 0.9cm]{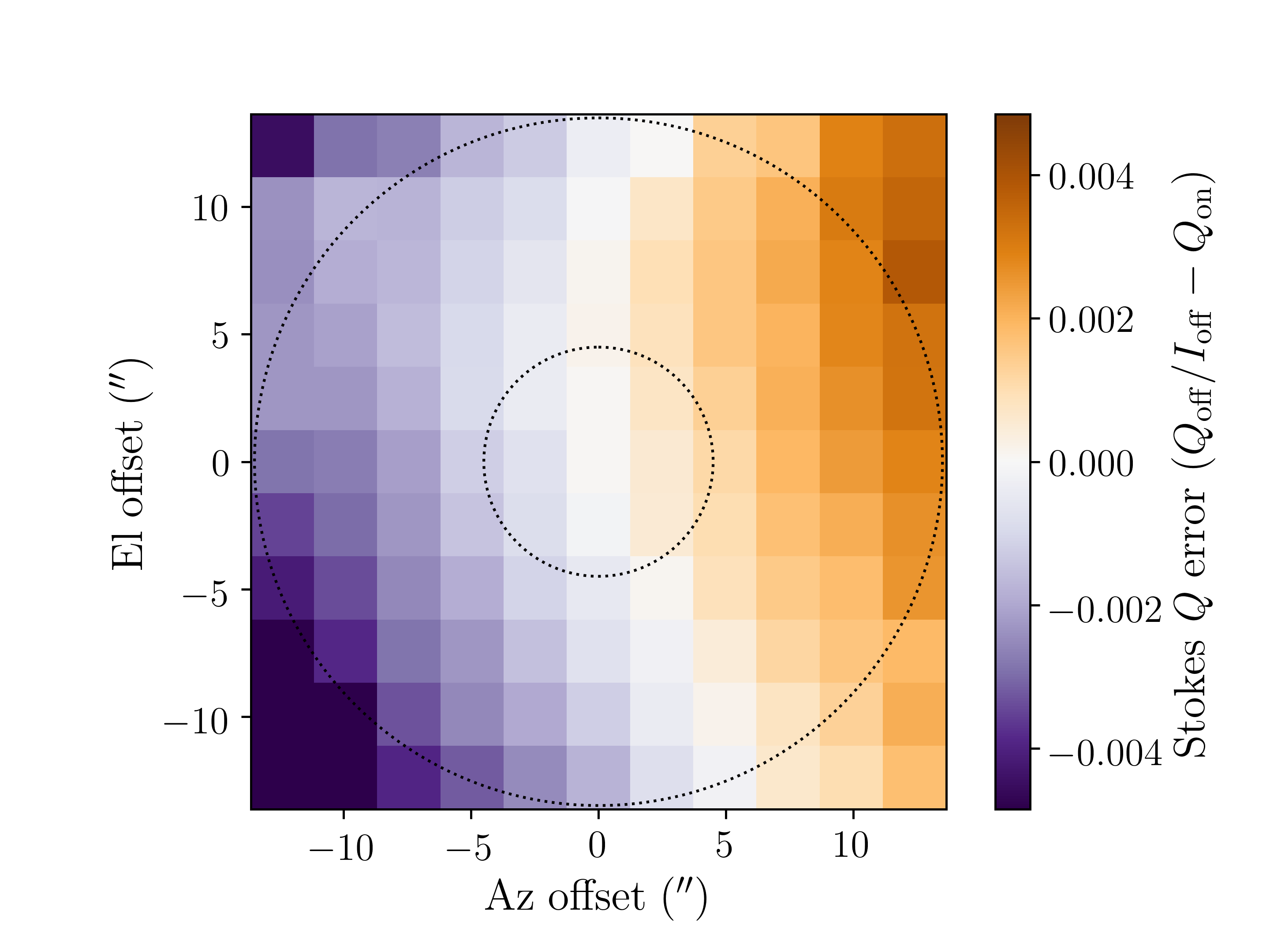}
\includegraphics[scale=0.58, clip, trim=0.9cm 0cm 0cm 0.9cm]{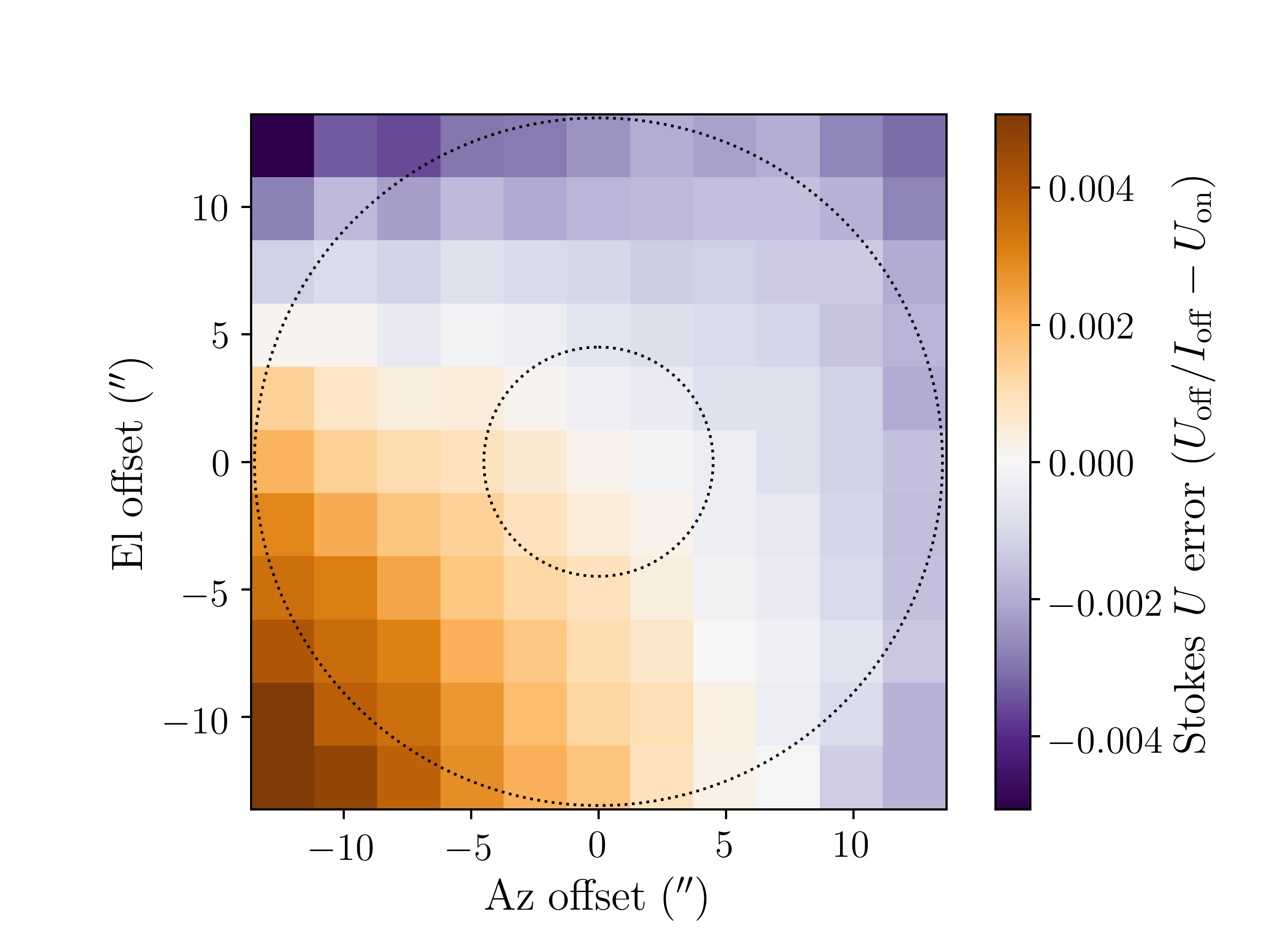} \\
\includegraphics[scale=0.58, clip, trim=0.9cm 0cm 0cm 0.9cm]{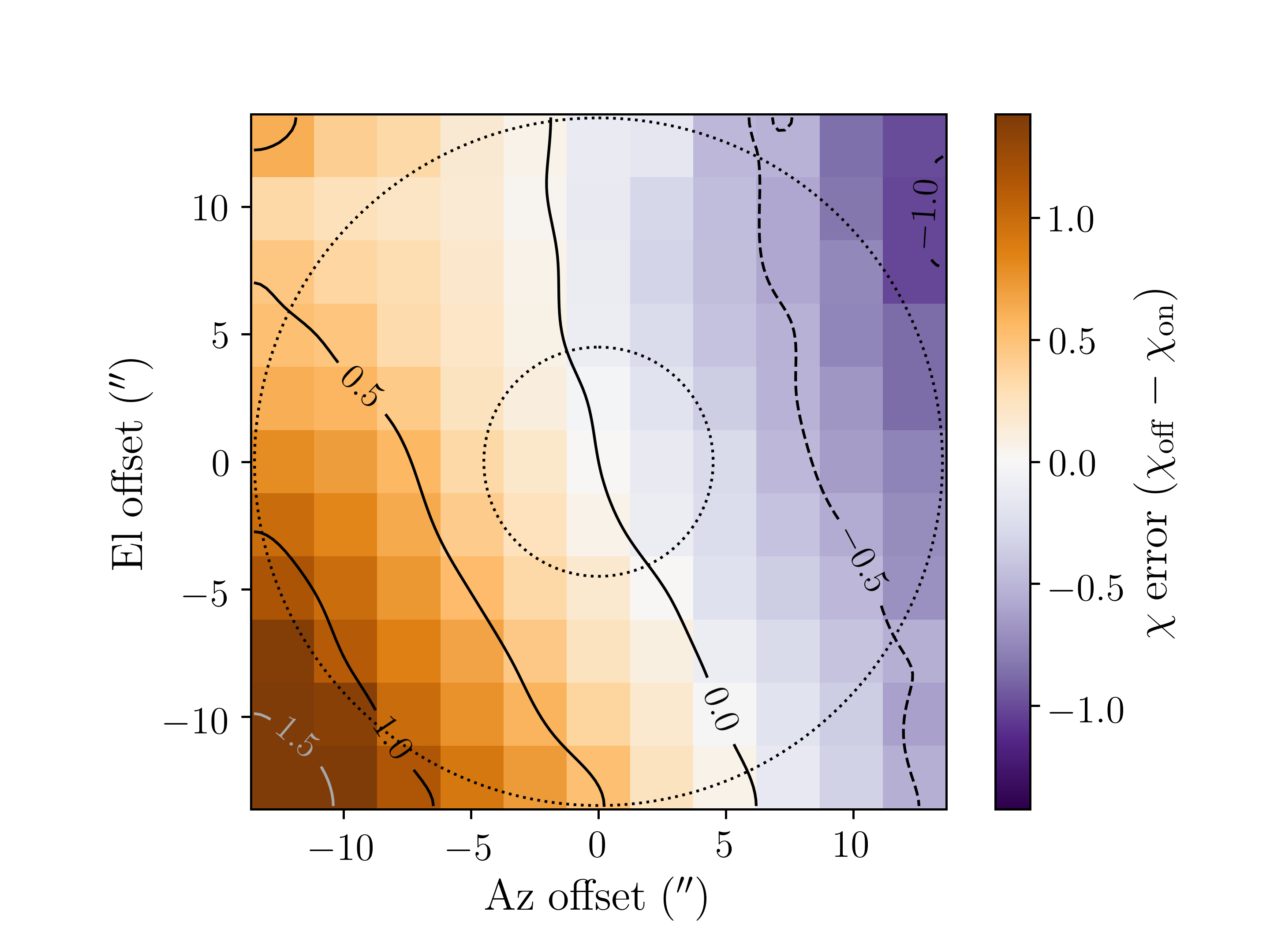}
\includegraphics[scale=0.58, clip, trim=0.9cm 0cm 0cm 0.9cm]{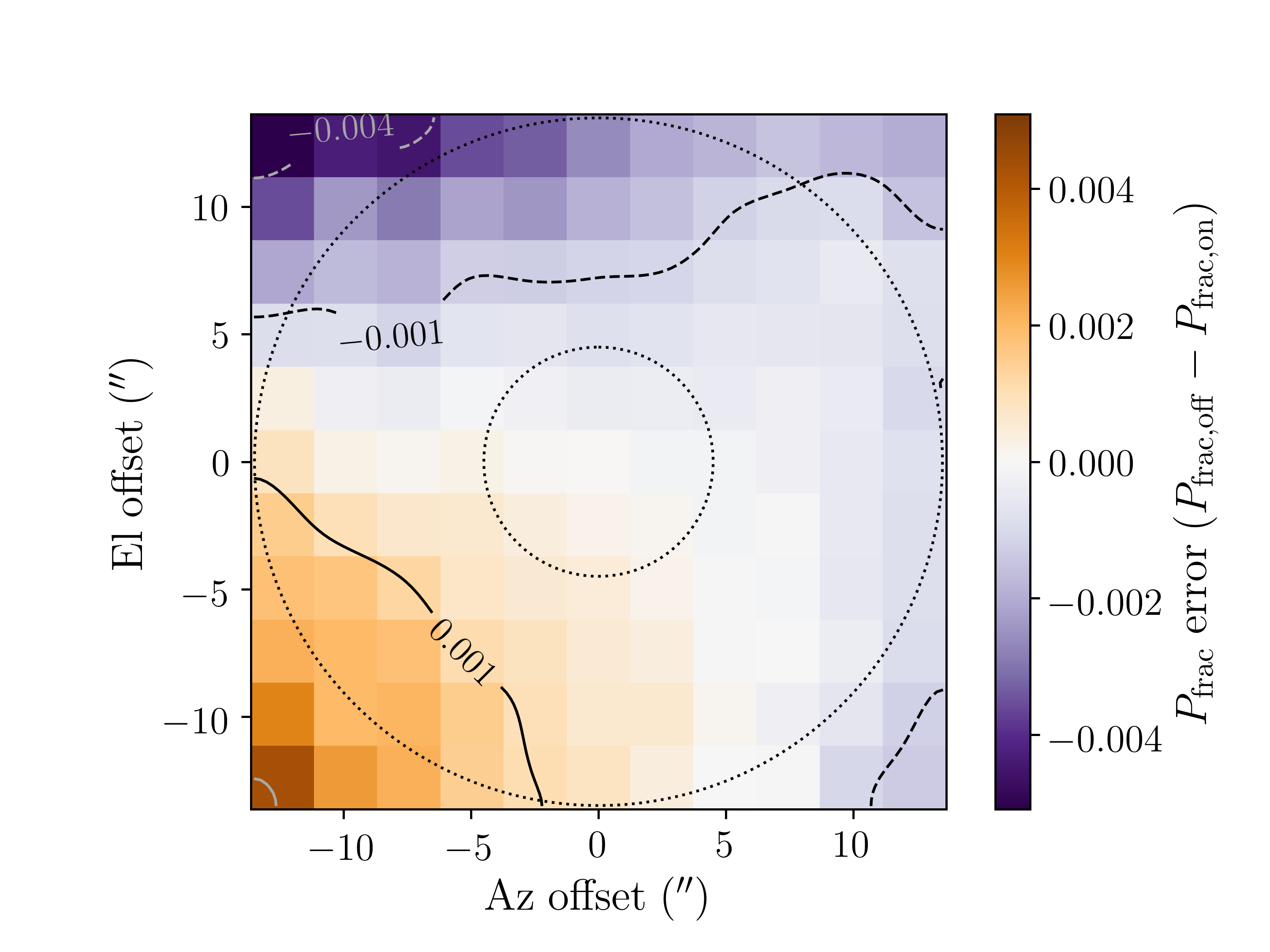}
\caption{\small Same as Figure \ref{fig:errors_B3-3}, for Band 6.}
\label{fig:errors_B6-1}
\end{figure*}

\begin{figure*} [hbt!]
\centering
\includegraphics[scale=0.58, clip, trim=0.9cm 0cm 0cm 0.9cm]{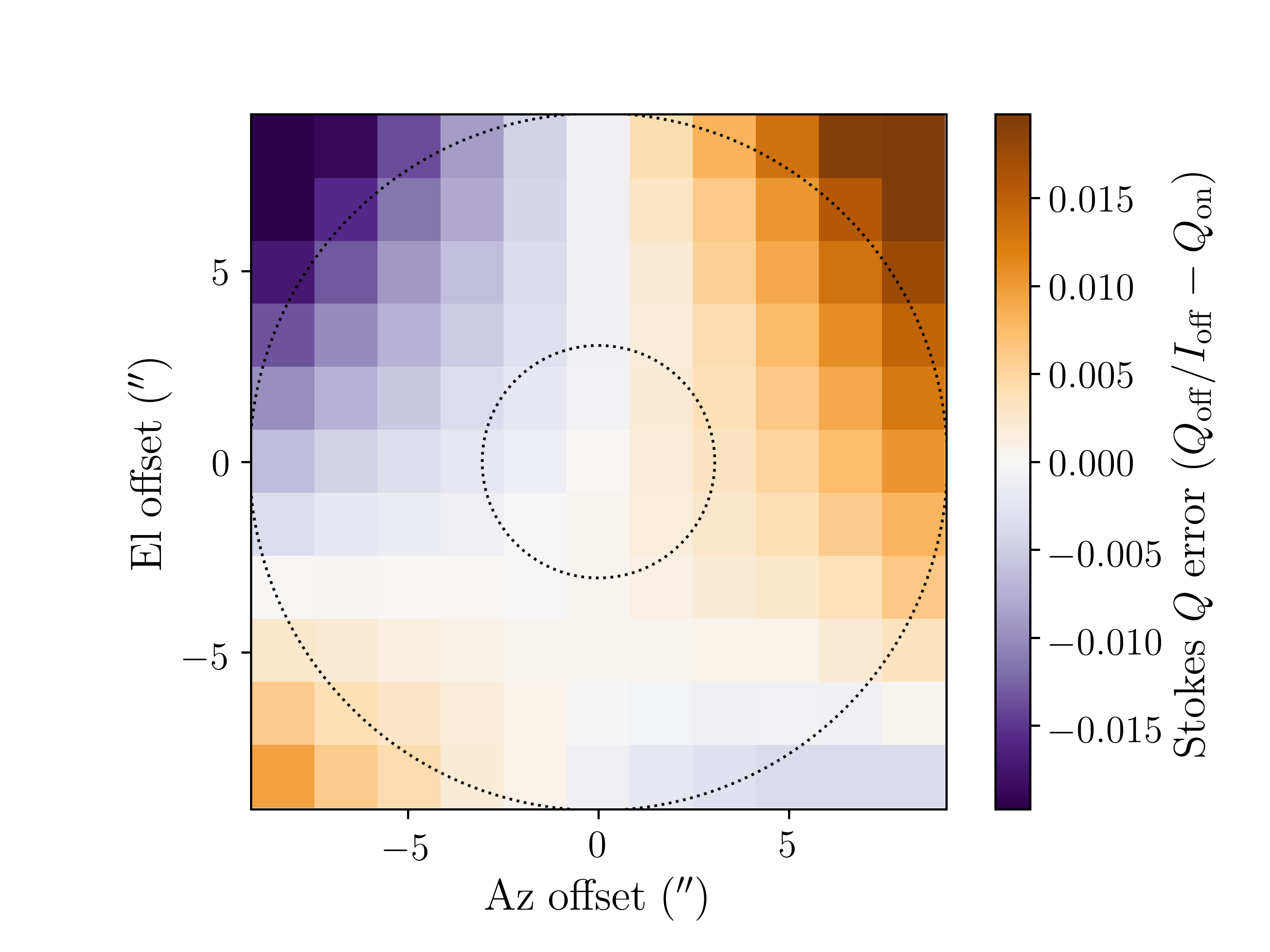}
\includegraphics[scale=0.58, clip, trim=0.9cm 0cm 0cm 0.9cm]{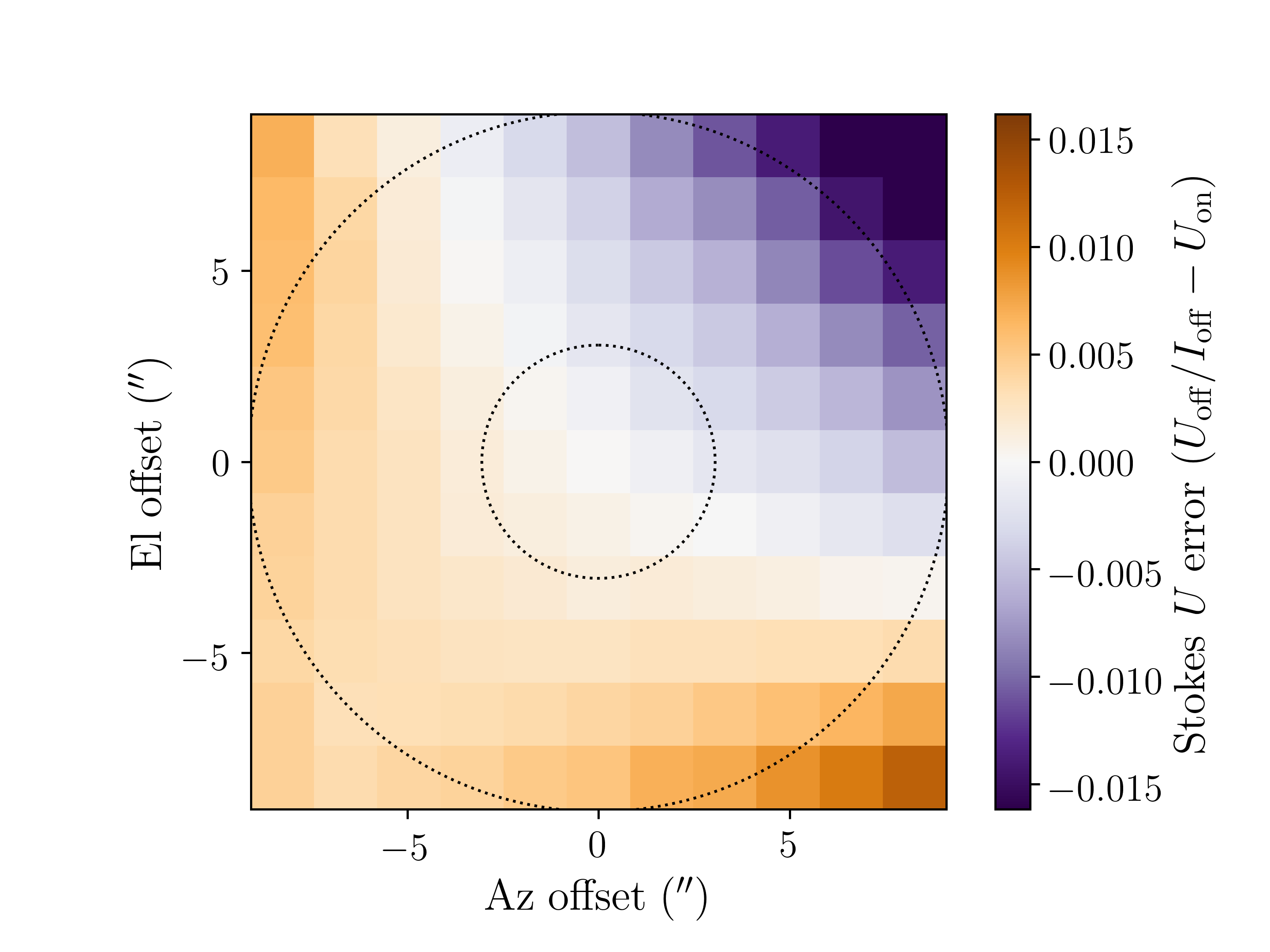} \\
\includegraphics[scale=0.58, clip, trim=0.9cm 0cm 0cm 0.9cm]{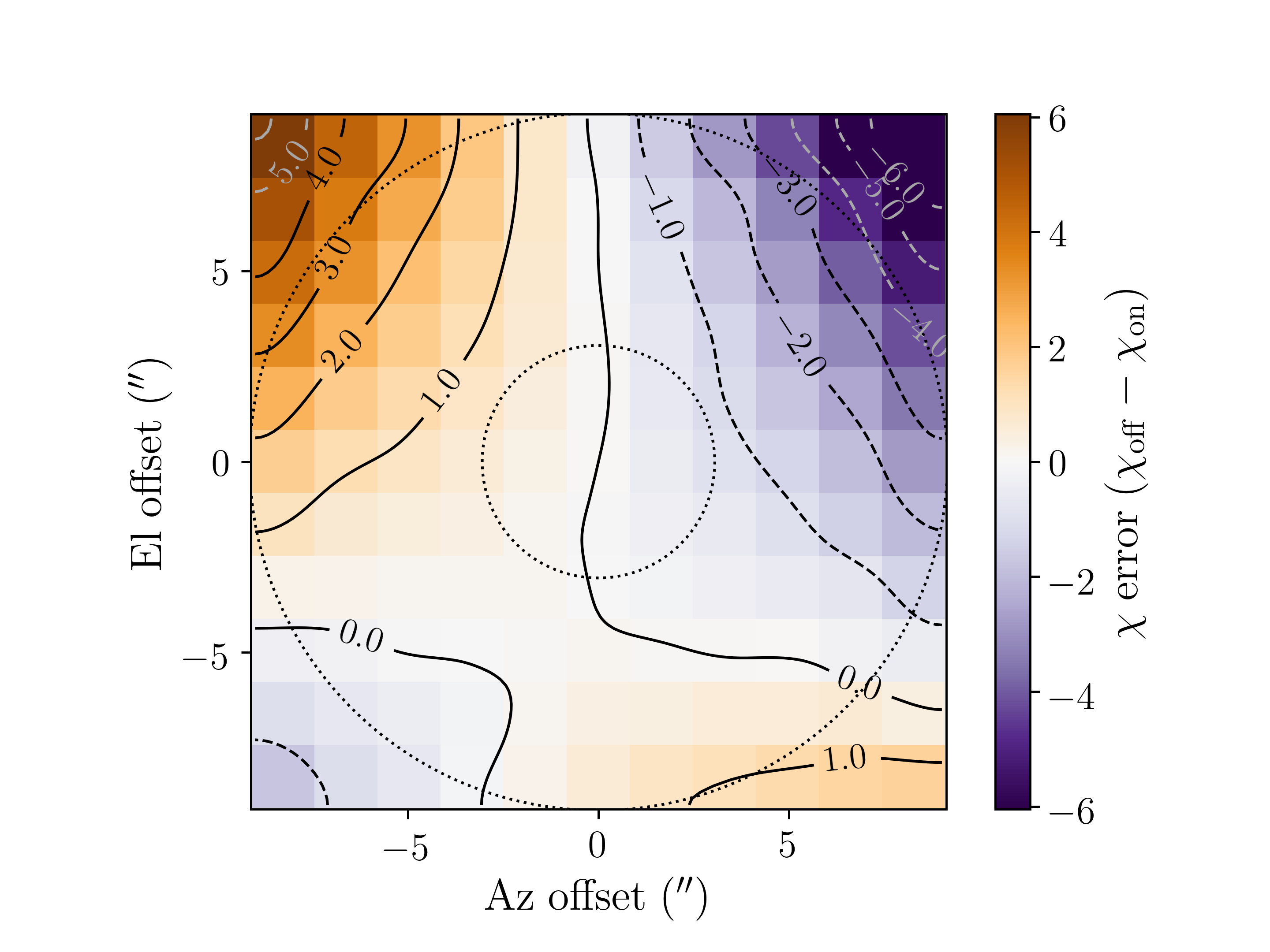}
\includegraphics[scale=0.58, clip, trim=0.9cm 0cm 0cm 0.9cm]{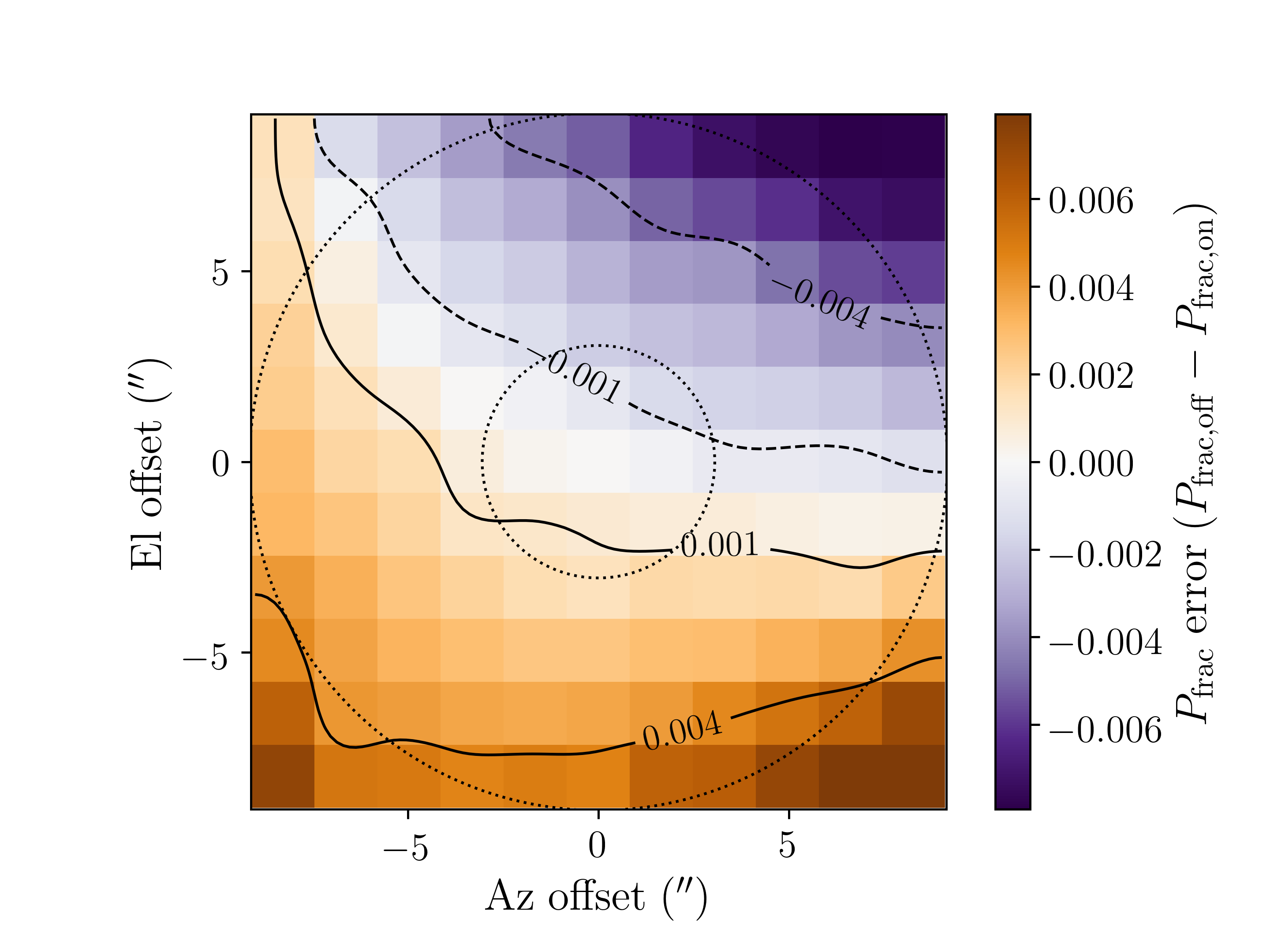}
\caption{\small Same as Figure \ref{fig:errors_B3-3}, for Band 7.}
\label{fig:errors_B7-1}
\end{figure*}

\clearpage


\subsection{\ref{app:error_11x11}. Error estimates in a polarization mosaic using the 11$\times$11 observations}
\label{app:error_11x11}

We perform a simple statistical analysis based on the the $11\times11$ observations of the highly linearly polarized blazar 3C~279 (see Sections \ref{sec:obs_11x11} and \ref{sec:results_11x11}) in order to use those data to estimate the errors in a polarization mosaic.  These estimates are an effort to connect the 11$\times$11 results with the error results from the Orion-KL mosaic data that we report in Section \ref{sec:results_orion}.

In single-pointing observations, the image can be divided into the inner region (which we define as $r\,<\,\frac{1}{3}$\,FWHM) and the outer ``donut'' region ($r > \frac{1}{3}$\,FWHM).  The initial ALMA polarization capabilities only allowed observations of objects whose emission falls in the inner region, as polarization errors are known to increase further off-axis, as shown in this paper.  However, in a mosaicked observation, neighboring pointings---each of which has varying polarization errors across its respective field of view---are stitched together, and thus the errors combine differently in different parts of the mosaic.  While much of a standard Nyquist-mosaicked image will fall within the $\frac{1}{3}$\,FWHM of a given pointing, there will still be a significant fraction of the image that will fall inside of the FWHM but outside of the $\frac{1}{3}$\,FWHM region of any given pointing: see the filled black region in Figure \ref{fig:mosaic_packing}.  In many cases, the wide-field polarization errors in multiple pointings will cancel out, thus reducing the overall error in the mosaicked map; however, the cancellation depends on a number of different factors, including the error patterns in the Stokes $Q$ and $U$ maps, the timing and duration of the observations, the chosen mosaic packing pattern, and the orientation of the source on the sky.  

In Section \ref{sec:results_11x11} we characterize the average polarized response of the primary beam of the 12\,m array at Bands 3, 5, 6, and 7 by observing a blazar in an 11$\times$11 grid in the Az/El frame.  We derive error maps in $Q$ and $U$ as well as in $\chi$ and $P_\textrm{frac}$.  Here we use those results to estimate the average error in the regions of the mosaic that are composed of overlapping outer regions of neighboring pointings.  In order to be conservative, in the analysis below we consider all of the data lying in the region $\frac{1}{3}$\,FWHM < $r$ < FWHM.  However, in practice, the data near the FWHM of a given pointing will be substantially down-weighted relative to the data from neighboring pointings, which will be closer to on-axis, since the center of each pointing in a Nyquist mosaic is located at the FWHM of its neighboring pointings.  

\begin{figure} [hbt!]
\centering
\includegraphics[scale=0.30, clip, trim=0cm 0cm 0cm 0cm]{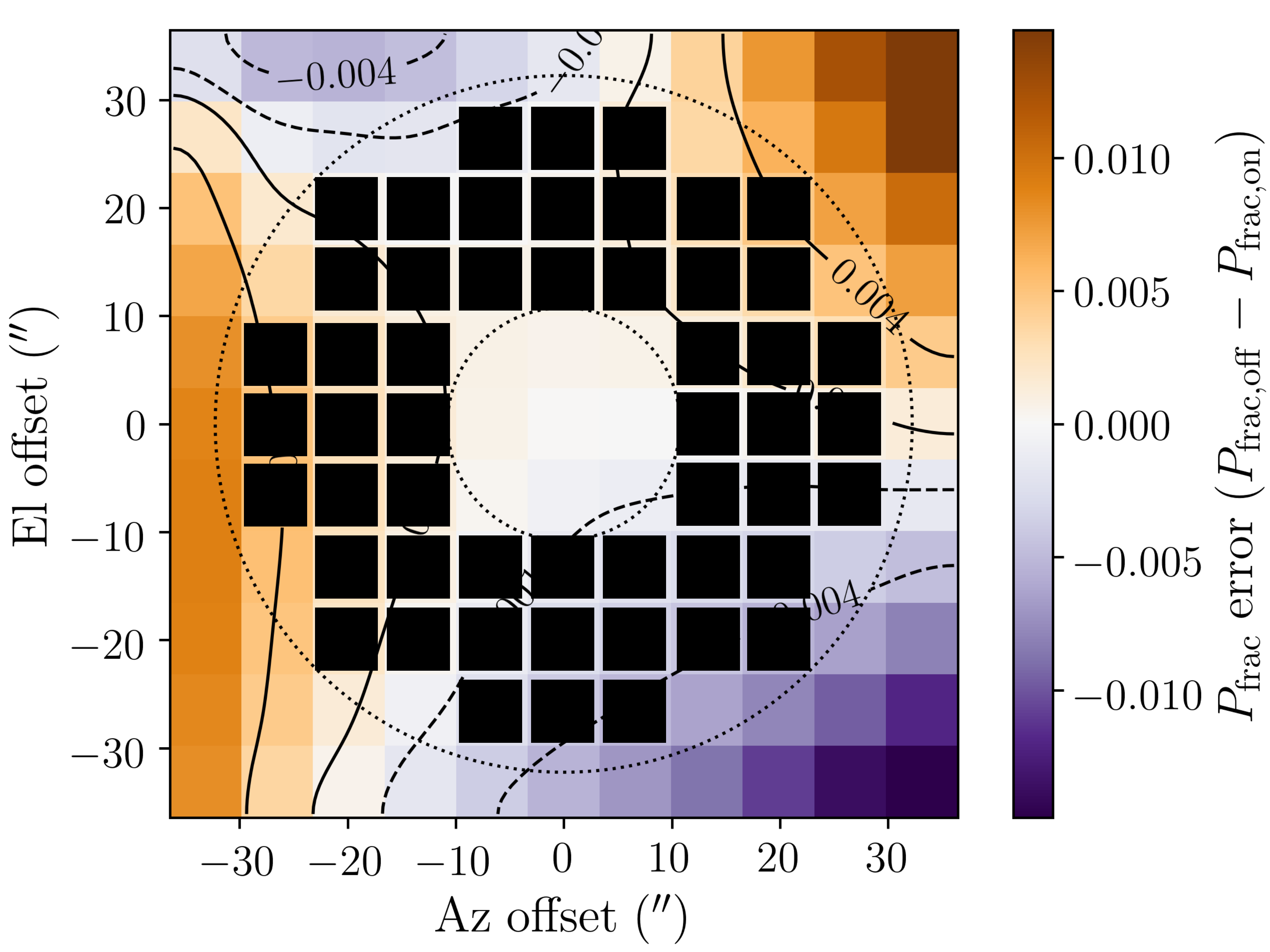}
\caption{\small The black squares indicate the pointings whose data were considered in the error estimates discussed in this Appendix.  As an illustration, they are overlaid on the $P_\textrm{frac}$ error map from the 11$\times$11 Band 3 test, whose original data are shown in Figure \ref{fig:errors_B3-3}.}
\label{fig:inner_ring}
\bigskip
\end{figure}

In order to estimate the error in the regions of a mosaic where multiple pointings overlap, we first use standard error propagation of the values in the maps of $\delta Q$ and $\delta U$.  These values, $\sigma_Q$ and $\sigma_U$, are calculated simply as the standard deviation of the values of $\delta Q$ and $\delta U$ in the ring that lies outside of the $\frac{1}{3}$\,FWHM, but inside the FWHM of a given pointing (see Figure \ref{fig:inner_ring} for a diagram of the chosen pointings). These $\sigma$ values correspond to estimators, in a statistical sense, of the ``typical'' errors in the off-axis $Q$ and $U$ values.  We then average $\sigma_Q$ and $\sigma_U$ to estimate the systematic error $\sigma_P$ in the polarized intensity $P$; this is justified since, almost always, $\sigma_Q \approx \sigma_U$ (see Table \ref{table:obs_orion}).  Thus, the error in the polarization fraction $\sigma_\textrm{Pfrac}$ can be approximated, using error propagation and the fact that $\sigma_Q \approx \sigma_U \equiv \sigma_P$, as:

\begin{align}
\sigma_\textrm{Pfrac} = P_\textrm{frac} \sqrt{ \left(\frac{\sigma_P}{P}\right)^2 + \left(\frac{\sigma_I}{I}\right)^2 } \,\,.
\end{align}

\noindent
Given that $I$ is large relative to $P$ and assuming a circular, azimuthally symmetric primary beam (and thus no position-dependent systematic error $\sigma_I$), we set the second term to zero.  This simplifies the equation to:

\begin{align}
\sigma_\textrm{Pfrac} = P_\textrm{frac} \frac{\sigma_P}{P}  \,\,,
\end{align}

\noindent
which is identically equal to the error in the polarized intensity $\sigma_P$, since $P_\textrm{frac} = P/I$ and we define $I$ to be 1.

The error in the polarization angle $\chi$ is:

\begin{align}
\sigma_\chi = \frac{0.5}{P^2} \sqrt{ \left( Q\,\sigma_U\right)^2 + \left(U\,\sigma_Q\right)^2 } \,\,.
\end{align}

\noindent
By again assuming that $\sigma_Q \approx \sigma_U \equiv \sigma_P$, we can simplify the equation to:

\begin{align}
\sigma_\chi = 0.5 \frac{\sigma_P}{P} \,\,.
\label{eqn:sigma_chi}
\end{align}

\noindent
Using these equations, we can estimate the typical error in $P_\textrm{frac}$ and $\chi$ in the areas indicated in Figure \ref{fig:inner_ring}.  See Table \ref{table:errors_from_11x11} for these values, which have superscript~$^a$.  

In addition to error propagation, another way to estimate the typical error in $P_\textrm{frac}$ and $\chi$ is to calculate directly the standard deviation of the values in the $\delta P_\textrm{frac}$ and $\delta \chi$ error maps. See Table \ref{table:errors_from_11x11}: these values, denoted by superscript~$^b$,  are the same as the values derived from error propagation to within factors of <\,40\% ($\sigma_\textrm{Pfrac}$) and <\,30\% ($\sigma_\chi$).  As mentioned above, the resulting errors are conservatively large, as they are derived from the entire region between the $\frac{1}{3}$\,FWHM and the FWHM.

\begin{table*}[tbh!]
\centering
\caption{\normalsize Errors in 11$\times$11 maps toward highly linearly polarized 3C~279}
\normalsize
\begin{tabular}{llllllll}
\hline \noalign {\smallskip}
Band & $\sigma_Q$ & $\sigma_U$ & $\sigma_P$ & $\sigma_\textrm{Pfrac}^a$ & $\sigma_\chi^a$ & $\sigma_\textrm{Pfrac}^b$ & $\sigma_\chi^b$  \\
     & & & & & (deg) & & (deg)                \vspace{0.05in} \\
\hline \noalign {\smallskip}
3 & 0.0040 & 0.0032 & 0.0036 & 0.0036 & 0.90 & 0.0025 & 1.1 \\
5 & 0.0020 & 0.0013 & 0.0016 & 0.0016 & 0.43 & 0.0013 & 0.52 \\ 
6 & 0.0016 & 0.0012 & 0.0014 & 0.0014 & 0.33 & 0.00085 & 0.43 \\ 
7 & 0.0041 & 0.0032 & 0.0037 & 0.0037 & 0.86 & 0.0024 & 1.1  \\ \hline
\end{tabular}
\smallskip
\tablecomments{\small $\sigma_Q$ and $\sigma_U$ are estimates of ``typical'' errors in the 11$\times$11 $Q$ and $U$ maps, calculated as described in this Appendix.  $\sigma_P$ is the average of $\sigma_Q$ and $\sigma_U$. \\
$^a$ Errors in polarization fraction ($\sigma_\textrm{Pfrac}$) and polarization position angle ($\sigma_\chi$) derived from error propagation of $\sigma_Q$ and $\sigma_U$. \\
$^b$ Errors estimated by taking the standard deviation of the values in the maps of $\delta P_\textrm{frac}$ and $\delta \chi$.}
\label{table:errors_from_11x11}
\bigskip
\end{table*}

\end{document}